\documentclass[preprint]{imsart}

\usepackage{amsthm,amsmath,amsfonts}
\RequirePackage[colorlinks,citecolor=blue,urlcolor=blue]{hyperref}
\usepackage[round]{natbib}
\usepackage[english]{babel}
\usepackage{latexsym}
\usepackage{subfigure}
\usepackage{epsfig}
\usepackage{color,soul}
\usepackage{moreverb}
\usepackage{mathtools}
\usepackage{tikz, color}
\usepackage{enumerate,linegoal}
\usepackage{calc}
\usepackage{latexsym}
\usepackage{amsmath}
\usepackage{amsfonts}
\usepackage{amssymb}
\usepackage{bm}
\usepackage{psfrag}
\usepackage{graphicx}
\usepackage{epstopdf}
\usepackage{framed}
\usepackage{booktabs}
 \usepackage[flushleft]{threeparttable}

\usepackage{xfrac,array,booktabs}
\usepackage{verbatim}
\usepackage{enumitem}

\DeclareMathOperator*{\argmin}{arg\,min}

\usetikzlibrary{matrix}

\startlocaldefs

\newcommand{\rd}{\mathrm{d}}
\newcommand{\rds}{\,\rd}
\renewcommand{\eqref}[1]{(\ref{#1})}

\newcommand{\tp}{^{\top}}

\endlocaldefs

\begin{document}

\begin{frontmatter}

\title{Additive quantile regression for clustered data with an application to children's physical activity}
\runtitle{Additive quantile mixed models}

\begin{aug}
\author{\fnms{Marco} \snm{Geraci}\corref{}\thanksref{t1}\ead[label=e1]{geraci@mailbox.sc.edu}}

\thankstext{t1}{Department of Epidemiology and Biostatistics, Arnold School of Public Health, University of South Carolina, 915 Greene Street, Columbia SC 29209, USA. \printead{e1}}

\runauthor{M. Geraci}

\affiliation{University of South Carolina\thanksmark{t1}}

\end{aug}

\begin{abstract}
\quad Additive models are flexible regression tools that handle linear as well as nonlinear terms. The latter are typically modelled via smoothing splines. Additive mixed models extend additive models to include random terms when the data are sampled according to cluster designs (e.g., longitudinal). These models find applications in the study of phenomena like growth, certain disease mechanisms and energy consumption in humans, when repeated measurements are available. In this paper, we propose a novel additive mixed model for quantile regression. Our methods are motivated by an application to physical activity based on a dataset with more than half million accelerometer measurements in children of the UK Millennium Cohort Study. In a simulation study, we assess the proposed methods against existing alternatives.
\end{abstract}



\end{frontmatter}

\section{Introduction}\label{sec:1}
The goal of regression analysis is to model the distribution of an outcome as a function of one or more covariates. Mean regression is used to assess how the outcome changes on average when the covariates change, and often implies that the direction and strength of the statistical associations are the same for all individuals in a population. However, conditionally on their observed characteristics, subjects who rank below or above the average of the outcome distribution may respond differently to the same treatment or exposure. Evidence of heterogeneous effects across the outcome distribution have been found in countless research including the effect of smoking on weight in lighter or heavier infants \citep{abrevaya_2001, koenker_hallock, geraci_2016a}; the effect of succimer chelation on different levels of cadmium concentrations in children's blood \citep{cao_etal}; or the effect of sedentary behavior and food prices on different centiles of children's anthropometric variables \citep{espana_etal,sturm_datar}. These children may be at higher risks of morbidity and mortality than those who are at the center of the distribution.

By definition, mean effects average out stronger and weaker effects. The averaging may even cancel out symmetric effects of same magnitudes but opposite signs on the tails of the distribution. Quantile regression (QR) \citep{koenker1978} is a flexible statistical tool with a vast number of applications that complements mean regression. QR has become a successful analytic method in many fields of science because of its ability to draw inferences about individuals that rank below or above the population conditional mean. The ranking within the conditional distribution of the outcome can be considered as a natural index of individual latent characteristics which cause heterogeneity at the population level \citep{koenker_geling}. There is an increasingly wider acknowledgement of the importance of investigating sources of heterogeneity to quantify more accurately costs, benefits, and effectiveness of interventions or medical treatments, whether it be an after-school physical activity program, a health care reform, or a thrombolytic therapy \citep[see, for example,][]{austin,beets_2016a,beyerlein,ding,rehkopf,wei_2015,winkelmann}. QR is particularly suitable for this purpose as it yields inferences that are valid regardless of the true underlying distribution. Also, quantiles enjoy a number of properties \citep{gilchirst_2000}, including equivariance to monotone transformations and robustness to outliers.

In this paper we are concerned specifically with nonparametric quantile regression functions of continuous response variables when data arise from cluster designs. Our research is motivated by a study on daily and weekly physical activity patterns in school-aged children using high-frequency accelerometer data. In general, temporal (diurnal) trajectories of physical activity are characterized by strongly nonlinear patterns that require some degree of smoothing \citep{butte_etal,fan_etal,morris_etal,sera_etal_2011,sera_etal_2017}. On the other hand, some predictors of interest may simply have linear effects. If, in addition, data are collected longitudinally to examine weekly patterns, then the correlation at the individual level must be taken into account.

We propose novel additive quantile models that include linear terms, nonlinear terms, as well as random-effects terms which account for the clustering. Further, nonlinear terms are modelled nonparametrically using penalized splines and fitted via automatic scatterplot smoothing within a mixed model framework \citep{ruppert_etal}. In the next section, we briefly describe the data. In Section~\ref{sec:3}, we describe the state of the art in nonlinear quantile regression and highlight the differences between our proposal and existing, potentially competing approaches. In Section~\ref{sec:4}, we describe the methods and, briefly, their implementation in the \texttt{R} language \citep{R}, with further technical details provided in Appendix A. In Section~\ref{sec:5}, we carry out a simulation study to assess the performance of the proposed methods (with tables reported in Appendix B). The real data analysis is presented in Section~\ref{sec:6} while concluding remarks are given in Section~\ref{sec:7}.

\section{Sedentariness and physical activity in UK children}
\label{sec:2}

The benefits of regular physical activity on well-being and life expectancy as well as the detrimental health effects of sedentary behavior have been amply documented \citep[e.g., see][]{ekelund_etal,katz_etal,warburton2006}. Physical inactivity in England is estimated to cost more than eight billion pounds a year. This includes both the direct costs of treating major, lifestyle-related diseases and the indirect costs of sickness absence \citep{nice_2008}. It is also estimated that 54,000 premature deaths a year are linked to a sedentary lifestyle \citep{dept_2002}.

Since establishing an active lifestyle at an early age is an important form of prevention against morbidity and premature mortality, promotion of physical activity in children and young people has gained a strategically central place in the public health agenda \citep{nice_2009}. A nationally representative study \citep{griffiths2013a} showed that only half of UK seven-year-olds achieve recommended levels of physical activity, with girls far less active than boys. It is therefore important to identify the predictors of physical activity, not only at the average intensity of activity but also (and perhaps especially) at lower and upper intensities.

\begin{figure}[t!]
\centering
\includegraphics[scale = 0.4]{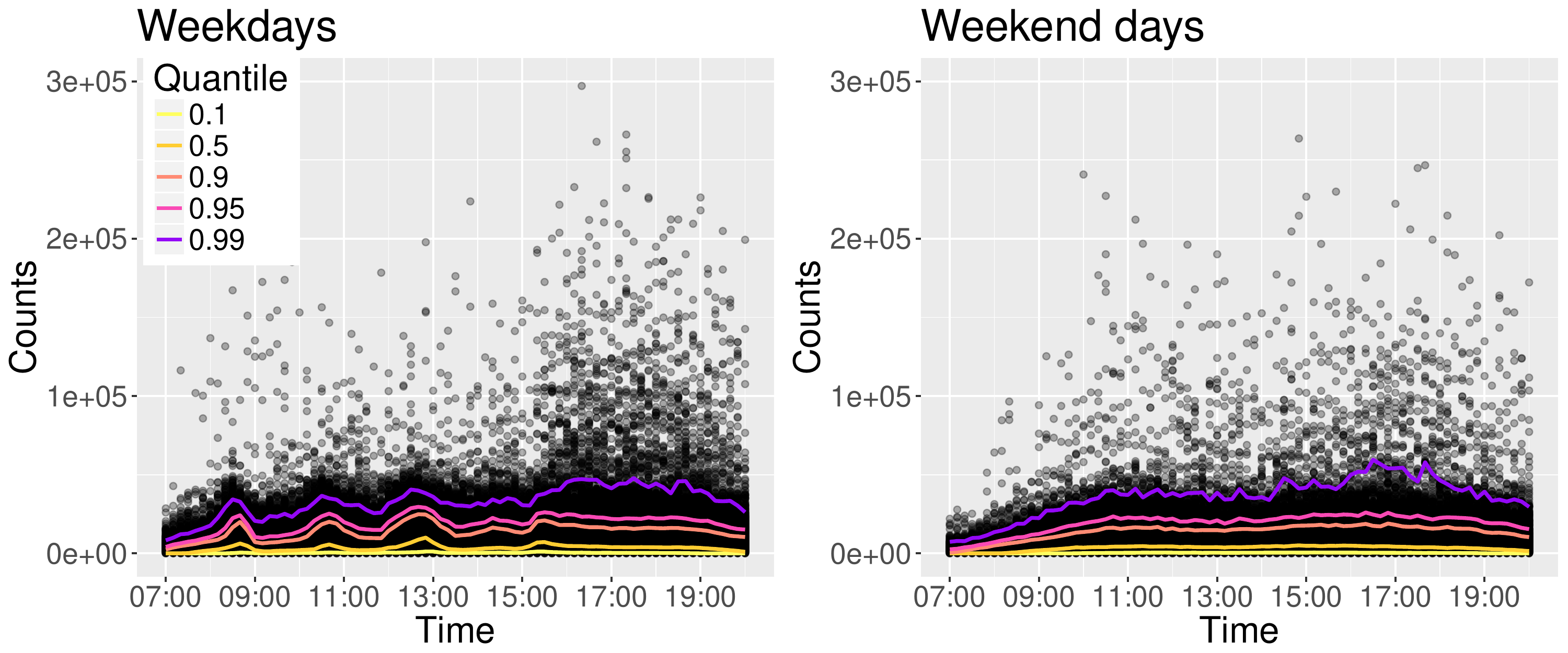}
\caption{Accelerometer counts (rounded to the nearest 100) observed between 7:00 and 20:00 and aggregated over 10-minute intervals in 1154 English children of the UK Millennium Cohort Study, by days of the week (Monday through Friday, weekdays; Saturday and Sunday, weekend). Solid lines connect sample quantiles that are estimated conditionally on time for 5 quantile levels $(0.1, 0.5, 0.9, 0.95, 0.99)$.}
\label{fig:1}
\end{figure}

Accelerometer data collected in seven-year-old children of the Millennium Cohort Study (MCS), a UK-wide longitudinal multi-purpose survey, represent a major, large-scale epidemiological resource to study physical activity determinants \citep{griffiths2013b}. Accelerometers are devices capable of providing an objective measure of the intensity and duration of movement. They produce an output known as `acceleration counts' which is dimensionless and thus requires calibration in order to be converted into physiologically more relevant units, usually based on energy expended per unit of time (e.g., metabolic equivalent of task). The plot in Figure~\ref{fig:1} shows accelerometer counts by time of the day for a subset of MCS children who provided reliable data for 7 days of the week. Since temporal trajectories of activity were similar between Monday and Friday, and during Saturday and Sunday, data in the plot are shown for weekdays (or workdays) and weekend days, respectively. Each dot represents the accelerometer counts measured on any of the weekdays or weekend days for a child at a given time of the day, while the solid lines are piecewise linear curves connecting sample quantiles estimated cross-sectionally (conditional on time). During weekdays there are periods of higher activity levels that mirror traveling times to and from school, and lunch and break times \citep{sera_etal_2017}. However, temporal trajectories at different quantile levels of the conditional distribution are not simply vertical shifts of one another. This suggests that the scale and possibly the shape of the counts distribution change with time of the day. For example, the skewness of the distribution in the weekend is small early in the morning, and steadily increases during the day.

\section{Approaches to nonlinear quantile regression}
\label{sec:3}
Nonlinear associations occur in many research studies, including bioassays and pharmacokinetic experiments \citep{lindsey}, as well studies related to growth processes in biology and agriculture \citep{davidian_2003}. In the presence of nonlinearity, there are different modelling strategies one may consider. Nonparametric models, smoothing splines (including polynomial models), and transformation models are the most commonly adopted. In addition to the specific strategy, one must also take into account the particular analytical framework: for example, is the sampling design cross-sectional or longitudinal? Is the outcome discrete or continuous? Should distributional assumptions be parametric or nonparametric? And so on. The effort needed for the analysis may have, at times, considerable weight on the final decision regarding which approach to follow. Lack of theory or even computer software can move the needle towards one choice over another.

The literature on parametric and nonparametric nonlinear regression models is vast. However, most of the work has been done in relation to nonlinear mean regression. In comparison, substantially less methods have been developed for nonlinear quantile regression and even less for nonlinear quantile regression with clustered data. A brief account of strategies to nonlinear quantile regression modelling is given by \cite{geraci_2016b} and these consist in: (i) smoothing (nonparametric regression); (ii) nonlinear parametric models; and (iii) transformation models. In the next sections, we discuss methods that fall in the first group and that are of direct relevance to the present study, and briefly survey those in the other two groups for completeness.

\subsection{Nonparametric models}
\label{sec:3.1}
Nonparametric regression and, more in general, additive models occupy an important place in statistical modelling. Popular approaches include locally weighted scatterplot smoothing \citep{cleveland1979,cleveland1988} and smoothing splines \citep{wahba1990, deboor2001, hastie1990, ruppert_etal}. If we consider the literature on nonparametric quantile functions for data with no clustering, there is a considerable variety of proposals. For example, \cite{koenker_portnoy}, \cite{he_etal}, \cite{koenker_mizera}, and \cite{koenker2011} considered total variation roughness penalties for univariate, bivariate, and additive quantile smoothing splines. For smooth functions, the total variation penalty corresponds to the $L_1$ counterpart of the $L_2$ smoothing spline penalty and leads to an elegant, computationally convenient optimization problem which can be solved via linear programming \citep{koenker_mizera}. \cite{yu1998}, \cite{horowitz2005}, and \cite{spokoiny2013} proposed estimation methods based on kernel weighted and polynomial local linear fitting. Other, similar local fitting methods are given in \cite{wu_etal} and \cite{cai2008}. In the Bayesian framework for independent data, \cite{thompson2010} proposed an approach based on natural cubic splines \citep{green1994} with a Gaussian smoothness prior.

A few approaches have been proposed also for the estimation of nonparametric quantile functions with repeated measurements or when the data are subject to other forms of dependence. \cite{wei_2006} considered modelling and estimation of longitudinal growth curves. They discussed an additive quantile regression model decomposed into a nonparametric temporal trend, a first-order autoregressive component, and a partially linear component to adjust for other covariates. \cite{fenske2013} extended \citeauthor{fenske2011}'s (\citeyear{fenske2011}) method to additive quantile regression based on longitudinal data. In their model, fitted via boosting, they included $L_{2}$-penalized fixed cluster-specific intercepts and slopes (thus, no covariance structure) with pre-specified shrinkage parameters, to account for individual deviations from a `population' trend. Additive nonlinear effects were modelled via penalized splines, again, with pre-specified amount of smoothing as controlled by the number of boosting iterations. An additive model is also considered by \cite{yue2011} who proposed normally distributed random intercepts and nonlinear terms with Bayesian P-splines and Gaussian Markov random fields as smoothness priors. Finally, it is worth mentioning that a related, although different, approach involves modelling the quantile regression coefficients in varying coefficient models, such as those by \cite{andriyana2014} and \cite{reich2011} for, respectively, longitudinally and spatially correlated data.

The modelling approach we develop in this study differs on several accounts from existing proposals. First of all, we model the intra-cluster correlation by means of random effects instead of autoregressive errors \citep{wei_2006}. Secondarily, in contrast to \cite{yue2011}, we include cluster-specific random slopes in addition to random intercepts, and, in contrast to \cite{fenske2013}, we allow cluster-specific effects to have a general covariance matrix. In addition, our estimation approach radically differs from that of \cite{fenske2013} since, as described in Section~\ref{sec:4}, the optimal degree of shrinkage of the cluster-specific effects and the optimal level of smoothing for the nonlinear terms are automatically estimated from the data.

\subsection{Nonlinear parametric models}
\label{sec:3.2}
Contributions to statistical methods for nonlinear mean regression when data are clustered can be found in the literature of mixed-effects modelling \citep{lindstrom_bates,pinheiro1995,pinheiro_bates} as well as generalized estimating equations \citep{davidian_1995,davidian_2003,contreras,vonesh_etal}. In contrast, there seem to be only a handful of published articles in the statistical literature on \emph{parametric} nonlinear quantile regression functions with clustered data. \cite{karlsson2008} considered nonlinear longitudinal data and proposed weighting the standard quantile regression estimator \citep{koenker1978} with pre-specified weights. \cite{wang2012}, taking her cue from \cite{geraci2007}, used the AL distribution to define the likelihood of a Bayesian nonlinear quantile regression model. \cite{huang2016} proposed a Bayesian joint model for time-to-event and longitudinal data. An approach based on copulas is developed by \cite{chen_etal}. \cite{oberhofer2016} established the consistency of the $L_{1}$-norm nonlinear quantile estimator under weak dependency. Finally, \cite{geraci_2017} extended \citeauthor{geraci2014}'s (\citeyear{geraci2014}) quantile mixed models to the nonlinear case.

\subsection{Transformation models}
\label{sec:3.3}
Traditionally, transformations toward linearity have been developed for conditional mean function estimation \citep[e.g.][]{aranda,box_cox}, with some proposals to deal with clustered data \citep[see, for example,][]{gurka_etal,hutmacher_etal,maruo_etal}. This general approach has been taken by others in quantile function estimation when data are independent using the Box-Cox transformation \citep{buchinsky,chamberlain,mu_he}, the Aranda-Ordaz transformation \citep{dehbi_etal}, as well as other new flexible transformations \citep{geraci_2015}. In particular, \cite{mu2009} developed a Box-Cox quantile regression model with varying coefficients for longitudinal data.

\section{Methods}\label{sec:4}

\subsection{Notation}
\label{sec:4.1}
We consider data from two-level nested designs in the form $(\mathbf{x}_{ij}\tp,\mathbf{z}_{ij}\tp,y_{ij})$, for $j=1,\ldots , n_{i}$ and $i=1,\ldots , M$, $N = \sum_i n_i$, where $\mathbf{x}_{ij}\tp$ is the $j$th row of a known $n_{i}\times p$ matrix $\mathbf{X}_i$, $\mathbf{z}_{ij}\tp$ is the $j$th row of a known $n_{i}\times q$ matrix $\mathbf{Z}_i$ and $y_{ij}$ is the $j$th observation of the response vector $\mathbf{y}_i = (y_{11},\ldots,y_{1n_{i}})\tp$ for the $i$th unit or cluster. This kind of data arise from longitudinal studies and other cluster sampling designs (e.g., spatial cluster designs). Throughout the paper, the covariates $x$ and $z$ are assumed to be given and measured without error. The $n \times 1$ vector of zeros and ones will be denoted by, respectively, $\mathbf{0}_{n}$ and $\mathbf{1}_{n}$, the $n \times n$ identity matrix by $\mathbf{I}_{n}$, and the $m \times n$ matrix of zeros by $\mathbf{O}_{m \times n}$. Finally, the Kronecker product and the direct sum will be denoted by $\otimes$ and $\oplus$, respectively.

\subsection{The model}
\label{sec:4.2}
We define the following $\tau$th additive quantile regression model
\begin{equation}\label{eq:1}
Q_{y_{ij}|\mathbf{u}_{i}, \mathbf{x}_{i},\mathbf{z}_{i}}(\tau) = \beta_{\tau,0} + \sum_{k=1}^{p} g^{(k)}_{\tau}(x_{ijk}) + \mathbf{z}_{ij}\tp\mathbf{u}_{\tau,i}, \qquad j = 1, \ldots, n_{i}, \quad i = 1,\ldots,M,
\end{equation}
for $\tau \in (0,1)$, where $g^{(k)}_{\tau}$ is a $\tau$-specific, centered, twice-differentiable, smooth function of the $k$th component of $\mathbf{x}$. The $q \times 1$ vector $\mathbf{u}_{\tau,i}$ collects cluster-specific random effects associated with $\mathbf{z}_{ij}$ and its distribution is assumed to depend on a $\tau$-specific parameter (further details are provided in the next section).

Without loss of generality, let the components of $\mathbf{x} = \left(x_{1}, \ldots, x_{s}, x_{s + 1}, \ldots, x_{p}\right)\tp$ be ordered in such a way that the first $s$ terms of the summation in \eqref{eq:1} are nonlinear functions and the remaining $p - s$ are linear. To model nonlinear functions, we consider a spline model of the type
\begin{equation*}
g_{\tau}(x) \approx \sum_{h = 1}^{H} v_{\tau,h}B_{h}(x),
\end{equation*}
(e.g., cubic or B-Spline), where the $B_{h}$'s and $v_{\tau,h}$'s, $h = 1, \ldots, H$, denote, respectively, the basis functions and the corresponding coefficients, and $H$ depends on the degrees of freedom or the number of knots. Note that the coefficients are $\tau$-specific.
The quantile function in \eqref{eq:1} is then approximated by
\begin{equation}\label{eq:2}
Q^{\ast}_{y_{ij}|\mathbf{u}_{i},\mathbf{x}_{i},\mathbf{z}_{i}}(\tau) = \beta_{\tau,0} + \sum_{k = 1}^{s} \sum_{h = 1}^{H_{k}} v_{\tau,hk}B^{(k)}_{h}(x_{ijk}) + \sum_{k=s+1}^{p} \beta_{\tau,k}x_{ijk} + \mathbf{z}_{ij}\tp\mathbf{u}_{\tau,i}.
\end{equation}
Let $\mathbf{B}^{(k)}(x_{ijk})$ be the $H_{k} \times 1$ vector of values taken by the $k$th spline evaluated at $x_{ijk}$, $\mathbf{v}_{\tau,k} = \left(v_{\tau,1}, \ldots, v_{\tau,H_{k}}\right)\tp$ be the $H_{k} \times 1$ vector of spline coefficients for the $k$th covariate, and $H = \sum_{k} H_{k}$. Further, define $\mathbf{B}_{i}$ as the $n_{i} \times H$ matrix with rows $\left(\mathbf{B}^{(1)}(x_{ij1})\tp, \ldots, \mathbf{B}^{(s)}(x_{ijs})\tp\right)\tp$, $j = 1, \ldots, n_{i}$, and let $\mathbf{v}_{\tau} = \left(\mathbf{v}_{\tau,1}\tp,\ldots,\mathbf{v}_{\tau,s}\tp\right)\tp$. With a slight abuse of notation, we write \eqref{eq:2} for the $i$th cluster in matrix form as
\begin{equation}\label{eq:3}
Q^{\ast}_{\mathbf{y}_{i}|\mathbf{u}_{i},\mathbf{X}_{i},\mathbf{Z}_{i}}(\tau) = \mathbf{F}_{i}\bm\beta_{\tau}+ \mathbf{Z}_{i}\mathbf{u}_{\tau,i} + \mathbf{B}_{i}\mathbf{v}_{\tau}, \quad i = 1,\ldots,M,
\end{equation}
where $\mathbf{F}_{i}$ is the $n_{i} \times (p - s + 1)$ matrix with rows $(1, x_{ij(s+1)}, \ldots, x_{ijp})\tp$, $j = 1,\ldots,n_{i}$, and $\bm\beta_{\tau} = \left(\beta_{\tau,0},\beta_{\tau,s+1}, \ldots, \beta_{\tau,p}\right)\tp$. We call \eqref{eq:3} an additive linear quantile mixed model, or AQMM for short.

The additive model introduced above opens up the question on how to control the trade-off between bias and efficiency, and, thus, the degree of smoothness of the estimate. At this juncture of our paper, we take a detour to briefly introduce the well-known link existing between penalized splines and mixed-effect models \citep[see, e.g.,][for an excellent review of this topic]{ruppert_etal}. Consider the following regression spline model
\begin{equation*}
\mathbf{y} = \mathbf{X}\bm\beta + \mathbf{B}\mathbf{v} + \bm\epsilon
\end{equation*}
for some $n \times p$ design matrix $\mathbf{X}$ and $n \times H$ basis matrix $\mathbf{B}$, where $\mathbf{y}$ is a vector of observations and $\bm\epsilon$ is a vector of independent and identically distributed (IID) normal errors $\mathcal{N}(0,\sigma^2)$ of dimension $n$. Penalized estimation of $\mathbf{v}$ can be carried out by minimizing the objective function
\begin{equation*}
\|\mathbf{y} - \mathbf{X}\bm\beta - \mathbf{B}\mathbf{v} \|^{2} + \lambda \| \mathbf{v}\|^{2}_{\mathbf{D}}
\end{equation*}
where $\lambda \geq 0$  is the \textit{smoothing parameter} and $\mathbf{D} = \mathbf{I}_{H}$. The above objective function, rescaled by $\sigma^2$, is equivalent to the best linear unbiased prediction criterion of the linear mixed-effects model $\mathbf{y} = \mathbf{X}\bm\beta + \mathbf{B}\mathbf{v} + \bm\epsilon$ with $\mathbf{v} \sim \mathcal{N}(\mathbf{0}, \sigma_{v}^2 \mathbf{I})$, $\sigma_{v}^2 = \sigma^2/\lambda$, and $\bm\epsilon \sim \mathcal{N}(\mathbf{0}, \sigma^2 \mathbf{I})$. Since the variance $\sigma_{v}^2$ is estimated from the data, it follows that the degree of smoothing is \textit{automatically} chosen by the estimation algorithm.

Automatic smoothing selection does not necessarily lead to optimal smoothing \citep{ruppert_etal}. However, one of the advantages of working with random spline coefficients when modelling cluster data is that $\mathbf{v}$ can be subsumed in the random part of the model containing the cluster-specific effects. Choice of the `prior' distribution for $\mathbf{v}$ effectively corresponds to choosing the form of the penalty. One approach is to use the same metric for the penalty term as that for the fidelity term. The $L_{1}$-penalty, which is linked to the double exponential distribution \citep{geraci2007,geraci2014}, is sometimes used in quantile regression models due to its computational convenience \citep{koenker_portnoy,koenker_mizera}. The resulting smoothed curves are piecewise linear and are most useful in the presence of breakpoints, sharp bends, and spikes \citep{koenker_mizera}.

In contrast, the $L_{2}$-penalty represents a more suitable choice for modelling smooth changes as in, for example, variations of energy expenditure over time. This is, for example, the approach considered by Cox and Jones in the discussion of \cite{cole} who suggested the spline smoothing quantile regression model
\begin{equation*}
\rho_\tau(y - f(x)) + \lambda \int \left\{f''(x)\right\}^{2} \rds x,
\end{equation*}
where $\rho_\tau(r)= r\left\{\tau-I(r < 0)\right\}$ is the quantile regression check function \citep{koenker1978} and $I$ denotes the indicator function. As compared to other roughness functionals, this kind of penalty yields a more visually appealing form of smoothness.

A natural link between $L_{2}$-penalized splines and random effects is provided by the normal distribution. Hence, in our random-effects specification of \eqref{eq:3}, we assume that the vectors $\mathbf{u}_{\tau,i}$ and $\mathbf{v}_{\tau}$ follow zero-centered multivariate Gaussians with variance-covariance matrices $\bm\Sigma_{\tau}$ and $\bm\Phi_{\tau}=\bigoplus_{k=1}^{s} \phi_{\tau,k}\mathbf{I}_{H_{k}}$, respectively. Further, we assume that the $\mathbf{u}_{\tau,i}$'s are independent for different $i$ (but may have a general covariance structure) and are independent from $\mathbf{v}_{\tau}$. Our objective function is then given by
\begin{equation}\label{eq:4}
\sum_{i=1}^{M} \rho_\tau\left(\mathbf{y}_{i} - \mathbf{F}_{i}\bm\beta_{\tau} - \mathbf{Z}_{i}\mathbf{u}_{\tau,i} - \mathbf{B}_{i}\mathbf{v}_{\tau} \right) + \sum_{i=1}^{M} \|\mathbf{u}_{\tau,i}\|_{\bm \Sigma_{\tau}^{-1}}^{2} + \sum_{k = 1}^{s} \phi^{-1}_{\tau,k}\|\mathbf{v}_{\tau,k}\|^{2},
\end{equation}
with the convention that $\rho_\tau(\mathbf{r})=\sum_{j=1}^{n} r_{j}\left\{\tau-I(r_{j} < 0)\right\}$ for a vector $\mathbf{r} = \left(r_{1},\ldots,r_{n}\right)\tp$. Note that the $\phi_{\tau,k}$'s determine the amount of smoothing for the nonparametric terms.

\subsection{Inference}
\label{sec:4.3}
The minimization of \eqref{eq:4} is equivalent to fitting a linear quantile mixed model (LQMM) \citep{geraci2007,geraci2014} where the asymmetric Laplace density
\begin{equation*}
p(y) = \frac{\tau(1-\tau)}{\sigma_{\tau}}\exp\left\{-\frac{1}{\sigma_{\tau}}\rho_\tau\left(y -\mu_{\tau}\right)\right\},
\end{equation*}
with location $\mu_{\tau} \in \mathbb{R}$ and scale $\sigma_{\tau} > 0$, is employed as quasi-likelihood for the fidelity term.

Define $\mathbf{y} = \left(\mathbf{y}_{1}\tp,\ldots,\mathbf{y}_{M}\tp\right)\tp$ and $\mathbf{u}_{\tau} = \left(\mathbf{u}_{\tau,1}\tp,\ldots,\mathbf{u}_{\tau,M}\tp\right)\tp$. Let $\bm\theta_{\tau}\equiv\left(\bm\beta_{\tau}\tp,\bm\xi_{\tau}\tp,\log{\bm\phi_{\tau}}\tp\right)\tp \in \mathbb{R}^{p+m+1}$ denote the parameter of interest, where $\bm\xi_{\tau}$ is an unrestricted $m$-dimensional vector, $1 \leq m \leq q(q+1)/2$, of non-redundant parameters in $\bm\Sigma_{\tau}$ \citep[e.g., see][]{pinheiro1996} and $\bm\phi_{\tau} = \left(\phi_{\tau,1}, \phi_{\tau,2}, \ldots, \phi_{\tau,s}\right)\tp$. Our goal is to maximize the marginal log-likelihood
\begin{align}\label{eq:5}
\nonumber & \ell\left(\bm\theta_{\tau}; \mathbf{y}\right) = N\log\left\{\frac{\tau(1-\tau)}{\sigma_{\tau}}\right\} - \frac{M}{2} \log |\tilde{\bm\Sigma}_{\tau}| - \frac{1}{2} \log |\tilde{\bm\Phi}_{\tau}|\\
\nonumber & \quad + \log\int_{\mathbb{R}^{H}}\left\{\prod_{i = 1}^{M}\int_{\mathbb{R}^{q}}\frac{\exp\left[-\left\{2\rho_\tau\left(\mathbf{y}_{i} - \mathbf{F}_{i}\bm\beta_{\tau} - \mathbf{Z}_{i}\mathbf{u}_{\tau,i} - \mathbf{B}_{i}\mathbf{v}_{\tau}\right) + \mathbf{u}_{\tau,i}\tp \tilde{\bm{\Sigma}}_{\tau}^{-1}\mathbf{u}_{\tau,i}\right\}/2\sigma_{\tau}\right]}{(2\pi\sigma_{\tau})^{q/2}} \rds \mathbf{u}_{\tau,i}\right\} \\
& \quad \times \frac{\exp\left(-\frac{1}{2\sigma_{\tau}}\mathbf{v}_{\tau}\tp \tilde{\bm\Phi}^{-1}_{\tau} \mathbf{v}_{\tau}\right)}{(2\pi\sigma_{\tau})^{H/2}} \rds \mathbf{v}_{\tau},
\end{align}
where $\tilde{\bm\Sigma}_{\tau} = \bm\Sigma_{\tau}/\sigma_{\tau}$ and $\tilde{\bm\Phi}_{\tau} = \bm\Phi_{\tau}/\sigma_{\tau}$ are the scaled variance-covariance matrices of the random effects. Note that this is a three-level hierarchical model, with the innermost grouping factor represented by the clusters $i$ and the outermost factor represented by one single-level group (i.e., the entire sample). Despite the three levels, we define $\hat{Q}^{(0)}_{\mathbf{y}_{i}|\mathbf{u}_{i}=0,\mathbf{X}_{i},\mathbf{Z}_{i}}(\tau) = \mathbf{F}_{i}\hat{\bm\beta}_{\tau} + \mathbf{B}_{i}\hat{\mathbf{v}}_{\tau}$ as the predictions at level $0$ since the smooth terms originally `belong' to the fixed design matrix. Similarly, we define $\hat{Q}^{(1)}_{\mathbf{y}_{i}|\mathbf{u}_{i},\mathbf{X}_{i},\mathbf{Z}_{i}}(\tau) = \mathbf{F}_{i}\hat{\bm\beta}_{\tau}+ \mathbf{Z}_{i}\hat{\mathbf{u}}_{\tau,i} + \mathbf{B}_{i}\hat{\mathbf{v}}_{\tau}$ as the predictions at level $1$ (i.e., at the cluster level).

Estimation proceeds using a double approximation:
\begin{enumerate}
\item the loss function $\rho_{\tau}(r)$ is first smoothed at the kink $r = 0$;
\item the integral in \eqref{eq:5} is then solved using a Laplacian approximation for the (smoothed) loss function.
\end{enumerate}

As proposed by \cite{geraci_2017}, we consider the following smooth approximation \citep{madsen1993,chen2007}:
\begin{equation}\label{eq:6}
\kappa_{\omega,\tau}(r)=\begin{cases}
r(\tau-1)-\frac{1}{2}(\tau-1)^2\omega & \text{if $r \leq (\tau-1)\omega$},\\
\frac{1}{2\omega}r^2& \text{if $(\tau - 1)\omega\leq r \leq \tau
\omega$},\\
r\tau-\frac{1}{2}\tau^2\omega & \text{if $r \geq \tau\omega$},
\end{cases}
\end{equation}
where $r \in \mathbb{R}$ and $\omega > 0$ is a scalar ``tuning'' parameter. A similar approximation is given by \cite{muggeo_etal} who claimed that their method provides a better approximation than \citeauthor{chen2007}'s (\citeyear{chen2007}) algorithm. However, no analytical evidence was provided in their paper to support such a claim. This point might offer scope for additional investigation but, here, it represents a secondary issue and will not be discussed any further.

We then replace the function $\rho_{\tau}$ in \eqref{eq:5} with $\kappa_{\omega,\tau}$ to obtain a smoothed likelihood and apply a second-order Taylor expansion \citep{pinheiro_chao} to the resulting exponent. After some algebra, we obtain the following Laplacian approximation
\begin{align*}
\ell_{\mathrm{LA}}\left(\bm\theta_{\tau}; \mathbf{y},\hat{\mathbf{w}}_{\tau}\right) =  N\log\left\{\frac{\tau (1 - \tau)}{\sigma_{\tau}}\right\} -\frac{1}{2} \left(\log |\tilde{\bm\Psi}_{\tau}\ddot{\mathbf{H}}| + \sigma_{\tau}^{-1}h_{0}\right),
\end{align*}
where $\tilde{\bm\Psi}$ is the scaled variance-covariance matrix of $\mathbf{w}_{\tau} =  \left(\mathbf{u}_{\tau}\tp, \mathbf{v}_{\tau}\tp\right)\tp$, and $h_{0}$ and $\ddot{\mathbf{H}}$ are the terms of order, respectively, 0 and 2 of the above-mentioned Taylor expansion around the mode $\hat{\mathbf{w}}_{\tau}$ (see Appendix A for more details).

When using the asymmetric Laplace as pseudo-likelihood, inference should be confined to point estimation \citep[see for example][]{yang2016}. Standard errors of non-random parameters estimates can be calculated using block bootstrap \citep{efron_tibshirani}, although this increases the computational cost. Bootstrap confidence intervals have been shown to have good coverage in LQMMs \citep{geraci2014}. Given the relatively large size of the MCS dataset, for the analysis in Section~\ref{sec:6} we implemented an adaptation of the method by \cite{kleiner_2014}. The general idea is to perform a bootstrap on several subsets of the original data and then summarize measures of uncertainty from all subsets. This strategy, called `bag of little bootstraps' (BLB), greatly reduces the computing cost when the sample size is large \citep[see][for more details]{kleiner_2012,kleiner_2014}. The original method was developed for IID observations. Since we are dealing with clusters, we adapted the BLB approach as follows:
\begin{enumerate}
\item sample without replacement $s$ subsets of size $b < M$ from the pool of $M$ clusters (random partition);
\item for each of the $s$ subset, repeatedly ($R$ times) take a bootstrap sample of size $M$ and fit an AQMM for each replicate;
\item for each of the $s$ subsets, calculate the bootstrap variance;
\item as final estimate of the standard error, take the square root of the average of the $s$ variances in step 3.
\end{enumerate}

As explained by the authors, the advantage of the BLB approach as compared to traditional bootstrap lies in the smaller size of the subsets. Although the nominal bootstrap sample size is $M$, there are at most $b$ unique clusters in each subset. To obtain a bootstrap replicate, we only need a sample from a multinomial distribution with $M$ trials and uniform probability over $b$ possible events. Estimation proceeds with a weighted likelihood, where the cluster-specific weights are given by the multinomial counts.

\subsection{Implementation}
The methods described in this section were implemented as an add-on to the \texttt{R} package \texttt{lqmm} \citep{geraci_2014}. The add-on is currently available from the author's website (\url{https://marcogeraci.wordpress.com}) and will appear in a future release of the main package. The core function made use of routines available from the \texttt{mgcv} \citep{wood_2006a} and \texttt{nlme} \citep{nlme} packages using syntax and options (e.g., selection of spline models) that are familiar to the users of these packages.

\section{Simulation study}
\label{sec:5}
We ran a simulation study to assess the proposed methods. In our analysis, we considered the two most relevant alternatives for additive regression modelling. The first is based on additive mixed-effects regression (AMM), which is available in the \texttt{R} package \texttt{mgcv}. This package makes use of \texttt{nlme}'s fitting routines. The other is represented by \citeauthor{fenske2013}'s (\citeyear{fenske2013}) additive fixed-effects quantile regression (AFEQR) for longitudinal data, which is available in the \texttt{mboost} package \citep{mboost}. Since the former approach aims at modelling the conditional expectation of the outcome under the assumption of normal errors, AQMM should have an advantage over AMM when the true errors are non-normal and the location-shift hypothesis of the normal model is violated. On the other hand, AFEQR is directly comparable to AQMM since they both aim at the conditional quantiles of the outcome with no assumption about the error distribution. However, as noted in Section~\ref{sec:3.1}, there are two basic differences between these two quantile regression approaches since in AQMM: (i) the cluster-specific effects are assumed to be random as opposed to fixed, thus a covariance structure between effects can be introduced; and (ii) the level of smoothing of the nonparametric terms is automatically estimated from the data (as reciprocal of the variance components) as opposed to prior specification. These are not necessarily advantages (or disadvantages) but they do represent aspects to consider when choosing a strategy for modelling and estimation.

The data were generated according to the following model
\begin{align}\label{eq:7}
\nonumber y_{ij} = & \,\, \beta_{0} + \beta_{1}\sin(x_{ij,1}) + \frac{\beta_{2}}{1 + \exp\{-(x_{ij,2} - 0.5)/0.1\}}\\
 & + \beta_{3}x_{ij,3} + \beta_{4}x_{ij,4} + \mathbf{z}_{ij}\tp\mathbf{u} + (1 + \gamma x_{ij,3})\epsilon, \qquad j = 1, \ldots, n \quad i = 1,\ldots,M
\end{align}
where $\bm\beta = (1, 4, 15, 4, 3)\tp$, $x_{ij,1} \sim \mathcal{U}(0,4\pi)$, $x_{ij,2} \sim \mathcal{U}(0,1)$, $x_{ij,3} \sim \mathrm{Bin}(1, 0.3)$, $x_{ij,4} \sim \mathcal{N}(0,1)$, $\mathbf{z}_{ij} = (1, x_{ij,4})\tp$, $\mathbf{u} \sim \mathcal{N}(\mathbf{0}, \bm\Sigma)$, and
\begin{equation*}
\bm\Sigma =
\left(
  \begin{array}{cc}
    2 & 0.8 \\
    0.8 & 1 \\
  \end{array}
\right).
\end{equation*}

In one scenario, we set $\gamma = 0$ (homoscedastic), while in a separate scenario we set $\gamma = 1$ (heteroscedastic). Within these two scenarios, the error was generated according to either a standard normal, a Student's $t$-distribution with 3 degrees of freedom, or a $\chi^{2}$-distribution with 3 degrees of freedom. Thus, in total there were $2 \times 3 = 6$ different models. For each model, a balanced dataset was generated according to 6 sample size combinations of $n \in\{5,10\}$ and $M \in \{50,100,500\}$, yielding $6 \times 6 = 36$ simulation cases. Each case was replicated $R=500$ times.

For each replication, we fitted the AQMM defined in \eqref{eq:2} for $\tau \in \{0.1, 0.5, 0.95\}$ using a cubic spline for the nonlinear terms associated with $x_{ij,1}$ and $x_{ij,2}$. The model also included a random intercept and a random slope for $x_{ij,4}$ with a symmetric positive-definite covariance matrix. We followed the estimation algorithm described in Appendix A. We used a Nelder-Mead algorithm to maximize the approximated log-likelihood and a tolerance of $10^{-5}$ for the relative change of the log-likelihood as the stopping criterion. The modal random effects \eqref{eq:A.7} were estimated using a Broyden-Fletcher-Goldfarb-Shanno (BFGS) algorithm with gradient calculated as in \eqref{eq:A.5}. Since different strategies can be used to determine the starting values, we considered a na\"{i}ve and a model-based approach. In the former case, we used the least squares (LS) estimate for $\bm\beta_{\tau}$, the identity matrix for $\bm\Sigma_{\tau}$, the mean of the absolute LS residuals for $\sigma_{\tau}$, and half the standard deviation of $y$ for $\omega$. The random effects were all set equal to 0. In the latter case, we used parameter and random effects estimates from an AMM. All the results reported in this paper refer to the latter approach since it gave a superior performance.

The AFEQR models were fitted following \citeauthor{fenske2013}'s (\citeyear{fenske2013}) recommendations for the settings of the boosting algorithm, namely the maximum number of boosting iterations, the step length parameter $\nu \in (0,1]$, and the degrees of freedom of the base-learners. The first two parameters were set to 5000 and $0.1$, respectively, as determined by cross-validation (separately for the homoscedastic and the heteroscedastic scenarios). As explained by the authors, these two `hyper-parameters' control the shrinkage of the estimates. The larger the step length (or the smaller the number of iterations), the more biased and shrunken to zero the estimates will be. The number of degrees of freedom was set to 3 for each term of the boosting algorithm. Since the degree of smoothing is controlled by the number of boosting iterations, the final degree of smoothing at the end of the algorithm can still reach a higher order than that imposed by the initial degrees of freedom \citep{fenske2013}. We used the \texttt{mboost} package (v. 2.8-1) in the \texttt{R} environment for statistical computing and graphics \citep{R} (v. 3.4.2) on a desktop computer with a 3.60GHz quad core i7-4790 processor and 32 gigabytes of RAM.

As a measure of performance, we calculated the bias and the root mean squared error (RMSE) of the level-1 predicted quantile functions. The RMSE of the predictions is given by
\begin{equation*}
\frac{1}{N}\sum_{i = 1}^{M}\sum_{j = 1}^{n} \left\{\hat{Q}^{(1)}_{y_{ij}}(\tau) - Q_{y_{ij}}(\tau)\right\}^{2},
\end{equation*}
where $Q_{y_{ij}}(\tau)$ denotes the true $\tau$th conditional quantile function based on model \eqref{eq:7}. Analogously, we calculated the \emph{relative} bias and RMSE of the coefficients for the linear terms, namely $\beta_{3}$ and $\beta_4$. Note that these coefficients do not vary across quantiles in the homoscedastic scenario. In contrast, the value of the coefficient for $x_{ij,3}$ depends on $\tau$ in the heteroscedastic scenario. Its `true' value was determined empirically by fitting a linear quantile regression model \citep{koenker1978} with 4th degree polynomials on the nonlinear terms for $10,\!000$ samples of size $5,\!000$. Finally, we determined the proportion of negative residuals (PNR)
\begin{equation*}
\frac{1}{N}\sum_{i = 1}^{M}\sum_{j = 1}^{n} I\left\{y_{ij} - \hat{Q}^{(1)}_{y_{ij}}(\tau) < 0\right\},
\end{equation*}
which is expected to be approximately equal to $\tau$. All summary measures were averaged over the replications.

Given the large number of results of the simulation study, all the tables are reported in Appendix B and the results are summarized here as follows.

\begin{enumerate}
\item Prediction of conditional quantiles:
\begin{description}[font=\normalfont\itshape]
\item[Homoscedastic scenario] AQMM showed bias and RMSE lower than those of AFEQR consistently across all sample sizes for $\tau \in \{0.1,0.95\}$. At the median, AQMM and AFEQR gave similar results, both showing low bias and RMSE values. The RMSE of AMM was particularly large in the case with asymmetric errors, as one would expect. However, in the case with normal errors, it was comparable to that of AQMM and surprisingly larger than that of AFEQR. Across quantiles, PNR rates for AQMM (and AFEQR) were equal to the expected nominal $\tau$ or differed at most by one hundredth.
\item[Heteroscedastic scenario] AQMM still had a clear advantage over AFEQR except at the median, where again the two performed similarly, and at $\tau = 0.95$ under skewed errors. The performance of AMM relative to AQMM or AFEQR was comparable to that seen in the homoscedastic scenario.
\end{description}
\item Estimation of $\beta_{3}$:
\begin{description}[font=\normalfont\itshape]
\item[Homoscedastic scenario] AQMM performed well as compared to AFEQR in terms of both bias and RMSE consistently across all quantiles and sample sizes. In particular, the bias of AQMM was much below $1\%$ for the most part. In contrast, AFEQR's bias ranged from $4\%$ to $15\%$ and was more severe on the tails. At the median, AMM's performance was comparable to that of AQMM in terms of bias, but it was otherwise inferior in terms of RMSE.
\item[Heteroscedastic scenario] As compared to the homoscedastic scenario, AQMM and AFEQR showed larger biases on the tails, with percentages ranging from below $1\%$ to $6\%$ and from $7\%$ to $20\%$, respectively. AQMM had, as before, consistently lower bias than AFEQR as well as lower RMSE, though the latter was occasionally higher for AQMM at lower sample sizes. At the median, the performance of AMM in terms of bias was rather poor under the scenario with skewed errors, but reasonable under the two scenarios with symmetric errors. In contrast, AMM was less efficient than AQMM in all three error scenarios at all sample sizes.
\end{description}
\item Estimation of $\beta_{4}$:
\begin{description}[font=\normalfont\itshape]
\item[Homoscedastic scenario] AQMM outperformed AFEQR in terms of both bias and RMSE consistently across all sample sizes for $\tau \in \{0.1,0.5\}$. AFEQR's bias was, again, particularly high and around $15\%$ at $\tau = 0.1$. At $\tau = 0.95$, the bias of AQMM was larger than that seen for $\beta_{3}$ and somewhat erratic in the case with asymmetric errors, presumably due to the fact that the random component of the model includes $x_{4}$ and thus the estimation of $\beta_{4}$ is affected by the Laplacian approximation. In contrast, AFEQR gave larger RMSE values than AQMM but, occasionally, lower biases. At the median, AMM performed well in terms of bias in all scenarios but was less efficient than AQMM.
\item[Heteroscedastic scenario] The results were similar to those obtained in the heteroscedastic scenario.
\end{description}
\end{enumerate}

Finally, we provide a brief report on the computational performance and the sensitivity of the results to different starting values. AQMM reached convergence in $98.2\%$ of the replications. The median number of iterations to reach convergence for one model was 19 (min 5, max 293), while the median of the smoothing parameter $\omega$ at the last iteration was approximately $7.1 \times 10^{-5}$ (min $4.0 \times 10^{-84}$, max $1.1 \times 10^{-2}$).

When using the na\"{i}ve approach to determine the starting values for AQMM, the estimation algorithm converged $100\%$ of the times, and took on average less iterations and a shorter time to converge. The median number of iterations to reach convergence for one model was 13 (min 3, max 20), while the median of the smoothing parameter $\omega$ at the last iteration was approximately $4.0 \times 10^{-4}$ (min $4.1 \times 10^{-6}$, max $4.9 \times 10^{-1}$). However, the bias and RMSE were in general slightly higher than those reported in Supplementary Materials. Regardless, the conclusions reached about the performance of AQMM relative to AFEQR and AMM were the same as those discussed above.

\section{Data analysis}
\label{sec:6}
The MCS accelerometer data were collected between May 2008 and August 2009 from participating children of the fourth sweep of the parent longitudinal survey, which provided information on several covariates, including socio-demographic and behavioural variables. A number of cleaning and processing procedures were applied to the raw accelerometer data \citep{geraci2012, rich2014} using the \texttt{R} package \texttt{pawacc} \citep{pawacc}. Out of $12,\!625$ children participating in the study, approximately $6,\!500$ provided reliable data, the latter defined as data from accelerometers that were deemed to have been worn for at least two days, at least 10 hours each day \citep{rich2013, griffiths2013b}. However, for the purpose of our analysis, we retained observations only for those children with reliable data between 7:00 and 20:00 of each day of the week.

\begin{table}
\caption{Categorical and continuous variables for English children of the Millennium Cohort Study. The dataset consists of $638,\!162$ accelerometer measurements, aggregated over 10-minute intervals, from a total of 1154 children. Note that the reference categories are the modal categories.}
\begin{tabular}{p{4cm} p{4cm} p{2cm} p{3cm}}
  \toprule
  Variable & Levels & Children  ($\%$) & Measurements ($\%$)\\
  \midrule
   Sex    & Male (reference) & 614 $(53.2)$ \\
          & Female & 540 $(46.8)$ \\
  Ethnicity & White (reference) & 962 $(83.4)$\\
            & Other than white$^{a}$ & 192 $(16.6)$ \\
  Income quintile              & 1             & 117 $(10.1)$\\
                               & 2             & 170 $(14.7)$ \\
                               & 3             & 220 $(19.1)$\\
                               & 4             & 318 $(27.6)$\\
                               & 5 (reference) & 329 $(28.5)$\\
  Reading for pleasure & Often (reference) & 998 $(86.5)$ \\
                       & Not often$^{b}$ & 156 $(13.5)$ \\
  Transportation to/from  & Active (reference) & 604 $(52.3)$\\
  school                  & Passive & 550 $(47.7)$\\
  Number of cars or vans       & 0             & 65 $(5.6)$\\
  owned                        & 1             & 412 $(35.7)$ \\
                               & 2 (reference) & 620 $(53.7)$\\
                               & 3 or more     & 57 $(5.0)$\\
  Day of the week & Monday through Friday  & & $455,\!830$ $(71.4)$\\
          & (reference)            & &                    \\
         & Saturday or Sunday & & $182,\!332$ $(28.6)$\\
  Season & Autumn &  & $230,\!285$ $(36.1)$\\
         & Winter &  & $13,\!509$ $(2.1)$\\
         & Spring &  & $82,\!634$ $(12.9)$\\
         & Summer (reference) &  & $311,\!734$ $(48.9)$\\
  \midrule
  Variable & Unit & FNS$^{c}$ \\
  \midrule
  Time of the day & min & --- \\
  BMI & kg/m$^2$& $(11.2, 15.1, 16.1, 17.5, 32.6)$\\
  Accelerometer counts & --- & $(0, 0.9, 3.1, 7.5, 297.1)$\\
  ($\times 1000$)      &     &                            \\
  \bottomrule
\end{tabular}
\begin{tablenotes}
\item $^{a}$~Mixed ethnicity, Indian, Pakistani, Bangladeshi, and Black.
\item $^{b}$~Less than once or twice a week.
\item $^{c}$~Five-number-summary: minimum, three quartiles, and maximum.
\end{tablenotes}
\label{tab:1}
\end{table}

We considered several covariates. Linear terms pertaining to the socio-demographic domain were sex (binary, reference: male) and ethnic group (binary, reference: white) of the child, and OECD equivalized income quintiles (categorical, reference: fifth quintile). Linear terms pertaining to the behavioural domain were time spent reading for enjoyment (binary, reference: often), mode of transport to/from school (binary, reference: active), number of cars or vans owned (categorical, reference: two). Linear terms pertaining to the temporal domain were day of the week (binary, reference: weekday) and calendar season (categorical, reference: summer). Finally, we considered three nonparametric terms, one for time of the day on weekdays, one for time of the day on weekends, and one for body mass index (BMI). The outcome variable was accelerometer counts. The analysis was restricted to singletons born in England. This decision was motivated by the ethnic composition of the sample, consisting of almost all white children in Wales, Scotland and Northern Ireland. Since ethnicity is a strong predictor of physical activity \citep{griffiths2013a} and ethnicity is confounded with country, we removed children from Celtic countries. Further, we excluded 15 children with missing information on ethnicity and BMI.  A summary of the dataset is given in Table~\ref{tab:1}. Our sample comprised 1,154 children for whom accelerometer counts were aggregated over 10-minute intervals between 7:00 and 20:00 (thus producing 79 time points), for seven days of the week. In total, this gave $N = 638,\!162$ accelerometer measurements (that is, $n_{i} = 79 \times 7$, $i = 1, \ldots, 1154$).

Using a similar notation as in \eqref{eq:2}, the $\tau$th additive linear quantile regression model was specified as
\begin{align}\label{eq:8}
\nonumber Q^{\ast}_{y_{ij}|\mathbf{u}_{i}, \mathbf{x}_{i},\mathbf{z}_{i}}(\tau) = & \,\, \beta_{\tau,0} + \sum_{h = 1}^{H_{1}} v_{\tau,1}B^{(1)}_{h}(t_{j,0}) + \sum_{h = 1}^{H_{2}} v_{\tau,2}B^{(2)}_{h}(t_{j,1}) + \sum_{h = 1}^{H_{3}} v_{\tau,3}B^{(3)}_{h}(\mathrm{BMI}_{i})\\
\nonumber & + \beta_{\tau,1}\mathrm{sex}_{i,1} + \beta_{\tau,2}\mathrm{ethnicity}_{i,1} +
\beta_{\tau,3}\mathrm{income}_{i,1} + \beta_{\tau,4}\mathrm{income}_{i,2} +\\
\nonumber & + \beta_{\tau,5}\mathrm{income}_{i,3} + \beta_{\tau,6}\mathrm{income}_{i,4} +
\beta_{\tau,7}\mathrm{reading}_{i,1} +
\beta_{\tau,8}\mathrm{transport}_{i,1}\\
\nonumber & + \beta_{\tau,9}\mathrm{cars}_{i,0} + \beta_{\tau,10}\mathrm{cars}_{i,1} + \beta_{\tau,11}\mathrm{cars}_{i,3+} +
\beta_{\tau,12}\mathrm{weekend}_{i,1}\\
& + \beta_{\tau,13}\mathrm{autumn}_{i} +
\beta_{\tau,14}\mathrm{winter}_{i} + \beta_{\tau,15}\mathrm{spring}_{i} + \mathbf{z}_{ij}\tp\mathbf{u}_{\tau,i},
\end{align}
for $\tau \in \{0.1,0.5,0.9,0.95,0.99\}$. For fitting purposes, the outcome was scaled by $10^4$. The variables $t_{j,0}$ and $t_{j,1}$, $j = 1, \ldots, 79$, denote the time of the day for, respectively, weekdays and weekend days. Time was expressed as minutes divided by $60 \times 24$ (e.g., with 0.29 corresponding to 7:00 and 0.83 to 20:00) and then centered about its mid-value (0.56 corresponding to 13:30). Similarly, BMI was centered about its mode (15.5 kg/m$^2$). Given the large size of the dataset, smooth terms were modelled using low-rank thin plate splines \citep{wood_2003}, which have been shown to possess optimal properties both statistically and computationally. The random effects $\mathbf{u}_{i} = (u_{i,0},u_{i,1},u_{i,t_0},u_{i,t_1})\tp$ were assumed to follow a multivariate normal distribution with symmetric positive-definite variance-covariance matrix
\begin{equation*}
\bm\Sigma =
\left[
  \begin{array}{cccc}
    \sigma^{2}_{0} & \sigma_{0,1} & \sigma_{0,t_0} & \sigma_{0,t_1} \\
     & \sigma^{2}_{1} & \sigma_{1,t_0} & \sigma_{1,t_1} \\
     &  & \sigma^{2}_{t_0} & \sigma_{t_0,t_1} \\
     &  &  & \sigma^{2}_{t_1} \\
  \end{array}
\right].
\end{equation*}

The first two terms of \eqref{eq:8} can be interpreted as the $\tau$th time-specific quantile function of accelerometer counts on an summer weekday for a boy of white ethnicity with modal BMI living in a household in the highest income quintile and with two cars, who reads often (at least once or twice a week) and walks or bikes from/to school (as opposed to moving by car or bus), and whose temporal (linear) trajectory belongs to the zero (or modal) random-effect cluster.

We made an attempt to fit an analogous AMM to obtain starting values for AQMM. However, the function \texttt{gamm} failed due to insufficient memory. We also tried with a smaller subset of 200 children, but the \texttt{gamm} function failed with a convergence error. Given the satisfactory simulation results, we therefore used the na\"{i}ve approach described in Section~\ref{sec:5} to determine the starting values.

\begin{figure}[t!]
\centering
\includegraphics[scale = 0.4]{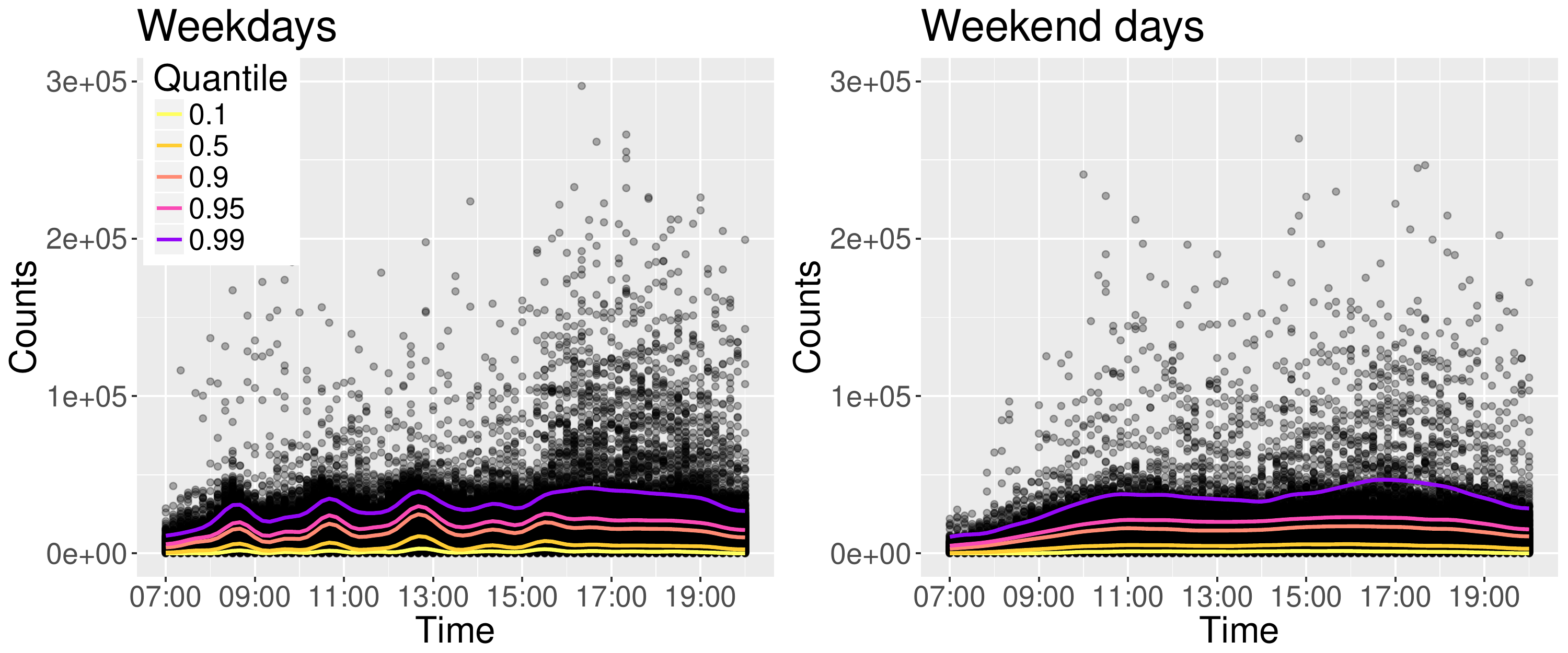}
\caption{Accelerometer counts observed between 7:00 and 20:00 and aggregated over 10-minute intervals in 1154 English children of the UK Millennium Cohort Study, by days of the week (Monday through Friday, weekdays; Saturday and Sunday, weekend). Solid lines represent conditional quantile functions estimated for a child in the reference group for 5 quantile levels $(0.1, 0.5, 0.9, 0.95, 0.99)$.}
\label{fig:2}
\end{figure}

The plot in Figure~\ref{fig:2} shows the estimated quantile function at level 0 for a child in the reference group. Diurnal patterns show markedly different shapes during the week. On weekdays, there are multiple peaks of activity in the morning and early afternoon, followed by a plateu of higher activity in the evening. On weekends, the trajectories look flatter and are characterized by two grand peaks around 11:00 and 17:00.

Estimates of the fixed effects and standard errors from AQMM are reported in Table~\ref{tab:2}. The latter were obtained using the BLB approach described in Section~\ref{sec:4.3} with a fivefold partition ($b \approx 230$) and $R = 50$ bootstrap replications. Some of the findings are consistent with those from previous analyses \citep{griffiths2013a,sera_etal_2017} that focused on the central part of the distribution, namely: girls and children of ethnicity other than white are less active than their peers; reading frequently during the week is negatively associated with activity; and higher activity levels characterize spring and summer, followed by autumn and winter.

However, the narrative emerging from Table~\ref{tab:2} is more variegated than this. First of all, there is a gradient across quantiles of increasingly larger differences in activity levels for girls and children of ethnicity other than white. Secondly, activity is lower in children from less affluent households at the most extreme quantile. In particular, activity is lower in those from economically disadvantaged (first quintile) across all quantiles. However, the estimates of the coefficients for income have large standard errors, resulting in statistical non-significance at the $95\%$ level. The effects associated with reading and mode of transportation does not seem to be important, neither practically nor statistically. In contrast, there are marked differences between children living in households with two vehicles (reference) and those with none, the latter being substantially more active. It also seems that at the quantile $0.99$, there is a $U$-shaped relationship between car/van ownership and activity counts.

While main effects of weekend on activity levels are approximately the same as those during the rest of the week across several quantiles, there is a rather strong positive weekend effect at the more extreme quantile. The results reported by \cite{sera_etal_2017} showed no weekend effect, which is likely the consequence of averaging out stronger and weaker effects. Finally, it interesting to note that the magnitude of the seasonal effects too increases with increasing quantiles. This is consistent with another quantile regression analysis of the MCS accelerometer data \citep{geraci_2016}.

\begin{figure}[t!]
\centering
\includegraphics[scale = 0.7]{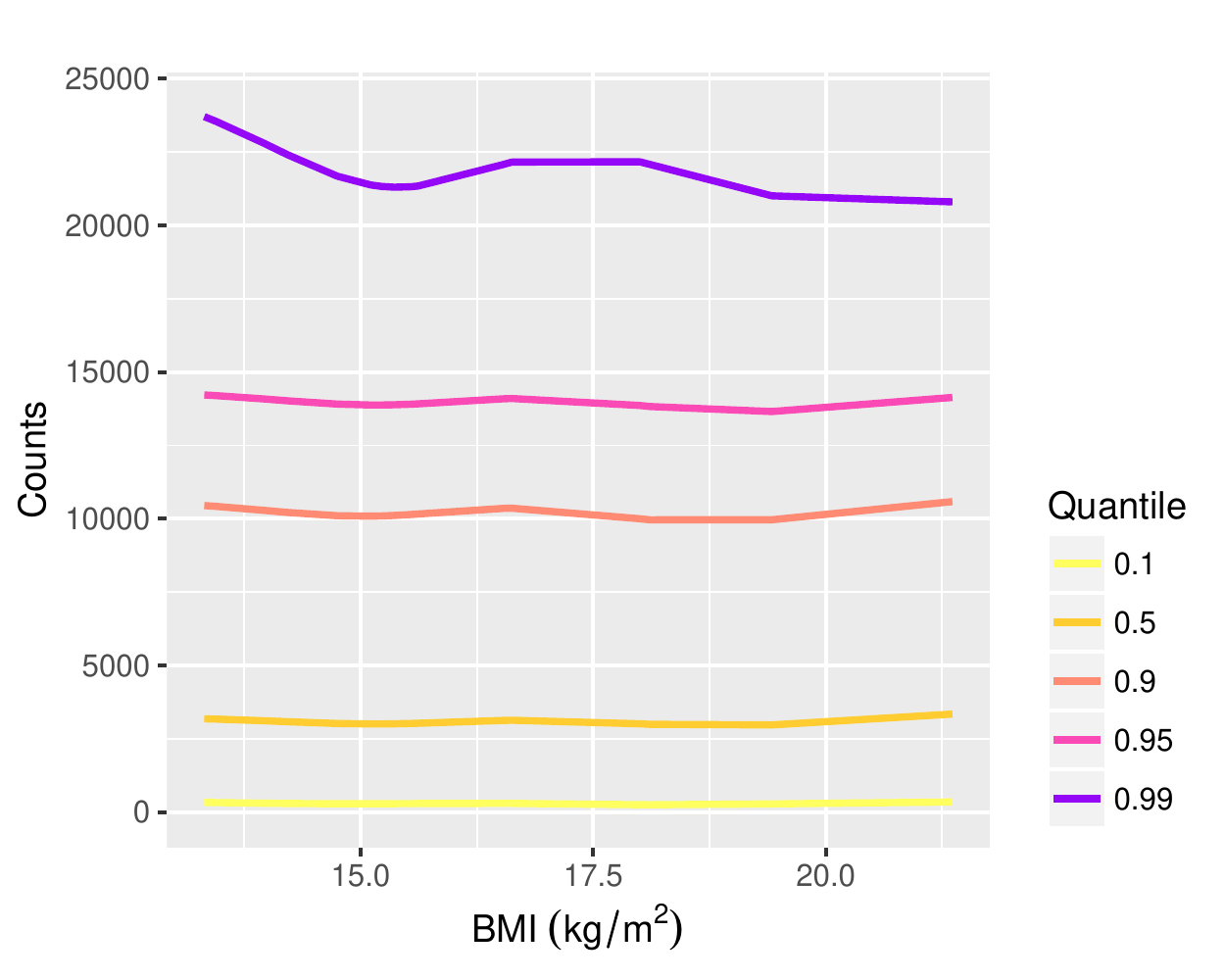}
\caption{Smooth functions of body mass index (BMI) estimated for a child in the reference group of the UK Millennium Cohort Study physical activity dataset for 5 quantile levels $(0.1, 0.5, 0.9, 0.95, 0.99)$.}
\label{fig:3}
\end{figure}

The estimated effect of BMI on activity counts for a child in the reference group is depicted in Figure~\ref{fig:3}. While the relationship is roughly constant up to the quantile $0.95$, it is nonlinear at $\tau = 0.99$, with an overall negative gradient. The variance of the corresponding smooth term (Table~\ref{tab:2}) indicates a stronger penalty on the spline coefficients at the most extreme quantile.

\begin{table}[ht]
\caption{Estimated fixed effects (counts per 10 minutes) and, in brackets, their standard errors, followed by estimated standard deviations and correlations of the random effects, standard deviations of the random spline coefficients, and proportion of negative residuals (PNR) from the additive quantile mixed model for the Millennium Cohort Study physical activity data. The reference categories are given in Table~\ref{tab:1}.}
\centering
\begin{tabular}{lrrrrr}
  \toprule
  \emph{Fixed effects} & \multicolumn{1}{c}{$\tau = 0.1$} & \multicolumn{1}{c}{$\tau = 0.5$} & \multicolumn{1}{c}{$\tau = 0.9$} & \multicolumn{1}{c}{$\tau = 0.95$} & \multicolumn{1}{c}{$\tau = 0.99$}\\
  \midrule
  Intercept & 992 (101) & 4408 (183) & 13704 (305) & 18473 (492) & 31065 (1136) \\
  Sex (female) & $-$24 (64) & $-$180 (95) & $-$2049 (222) & $-$2752 (328) & $-$3113 (843) \\
  Ethnicity (not white) & $-$101 (104) & $-$82 (124) & $-$1126 (285) & $-$1696 (388) & $-$3964 (948) \\
  Income quintile (1) & $-$43 (146) & $-$39 (216) & $-$483 (422) & $-$784 (567) & $-$2747 (1525) \\
  Income quintile (2) & 70 (124) & 99 (148) & 35 (337) & $-$2 (519) & $-$369 (1679) \\
  Income quintile (3) & 53 (95) & 13 (129) & $-$237 (303) & $-$512 (448) & $-$1196 (1009) \\
  Income quintile (4) & $-$44 (83) & 35 (129) & $-$135 (273) & $-$56 (412) & 776 (1089) \\
  Reading for pleasure (not often) & 92 (114) & 122 (141) & 8 (367) & 69 (502) & $-$276 (1179) \\
  Transportation (passive) & 62 (72) & 86 (87) & $-$274 (226) & $-$409 (363) & $-$209 (750) \\
  Number of cars or vans (0) & 25 (169) & $-$77 (224) & 1121 (549) & 1279 (581) & 3315 (739) \\
  Number of cars or vans (1) & 53 (83) & 75 (100) & 564 (231) & 682 (368) & 2083 (850) \\
  Number of cars or vans (3+) & 4 (164) & $-$53 (224) & 518 (496) & 655 (809) & 2586 (1072) \\
  Day of the week (weekend) & $-$148 (103) & $-$131 (106) & $-$168 (223) & $-$45 (364) & 3023 (1066) \\
  Season (autumn) & 12 (76) & $-$164 (91) & $-$958 (209) & $-$1067 (313) & $-$3155 (774) \\
  Season (winter) & $-$3 (199) & $-$204 (326) & $-$1377 (561) & $-$1675 (705) & $-$2999 (1124) \\
  Season (spring) & 82 (117) & 244 (156) & 1242 (333) & 2197 (528) & 6027 (2224) \\
  Linear basis term & 410 (50) & 782 (33) & 2050 (54) & 2639 (107) & 5374 (448) \\
  for time of the day (weekdays) & & & & \\
  Linear basis term & 635 (56) & 980 (46) & 2620 (83) & 3434 (158) & 7659 (965) \\
  for time of the day (weekend) & & & & \\
  Linear basis term for BMI & $-$40 (125) & $-$11 (51) & $-$16 (113) & $-$64 (166) & $-$95 (387) \\
  \midrule
  \emph{Standard deviations (random effects)} & \multicolumn{1}{c}{$\tau = 0.1$} & \multicolumn{1}{c}{$\tau = 0.5$} & \multicolumn{1}{c}{$\tau = 0.9$} & \multicolumn{1}{c}{$\tau = 0.95$} & \multicolumn{1}{c}{$\tau = 0.99$}\\
  \midrule
  $\hat{\sigma}_{0}$ (intercept weekdays)& 2969 & 3923 & 2882 & 2769 & 4897 \\
  $\hat{\sigma}_{1}$ (intercept weekend)& 3015 & 3526 & 2842 & 2809 & 5054 \\
  $\hat{\sigma}_{t_{0}}$ (time of the day weekdays)& 2868 & 3575 & 2817 & 2800 & 5069 \\
  $\hat{\sigma}_{t_{1}}$ (time of the day weekend)& 2867 & 3376 & 2940 & 2858 & 5017 \\
  \midrule
  \emph{Correlations (random effects)} & \multicolumn{1}{c}{$\tau = 0.1$} & \multicolumn{1}{c}{$\tau = 0.5$} & \multicolumn{1}{c}{$\tau = 0.9$} & \multicolumn{1}{c}{$\tau = 0.95$} & \multicolumn{1}{c}{$\tau = 0.99$}\\
  \midrule
  $\hat{\rho}_{0,1}$ & 0.93 & 0.97 & 0.73 & 0.36 & 0.52 \\
  $\hat{\rho}_{0,t_0}$ & 0.99 & 0.99 & 0.94 & 0.89 & 0.93 \\
  $\hat{\rho}_{1,t_0}$ & 0.93 & 0.97 & 0.71 & 0.35 & 0.51 \\
  $\hat{\rho}_{0,t_1}$ & 0.93 & 0.97 & 0.73 & 0.37 & 0.52 \\
  $\hat{\rho}_{1,t_1}$ & 0.99 & 0.99 & 0.93 & 0.88 & 0.93 \\
  $\hat{\rho}_{t_0,t_1}$ & 0.93 & 0.97 & 0.72 & 0.36 & 0.51 \\
  \midrule
  \emph{Standard deviations (smooth terms)} & \multicolumn{1}{c}{$\tau = 0.1$} & \multicolumn{1}{c}{$\tau = 0.5$} & \multicolumn{1}{c}{$\tau = 0.9$} & \multicolumn{1}{c}{$\tau = 0.95$} & \multicolumn{1}{c}{$\tau = 0.99$}\\
  \midrule
  $\hat{\phi}_{\mathrm{weekdays}}$ & 4136 & 15215 & 4114 & 4343 & 1722 \\
  $\hat{\phi}_{\mathrm{weekend}}$ & 8385 & 14541 & 6699 & 2402 & 515 \\
  $\hat{\phi}_{\mathrm{BMI}}$ & 2905 & 4945 & 7094 & 2777 & 181 \\
  \midrule
  \emph{PNR} & 0.11 & 0.50 & 0.90 & 0.95 & 0.99 \\
  \bottomrule
\end{tabular}
\label{tab:2}
\end{table}

The estimated standard deviations of the random effects show larger variability of individual linear trends (intercepts and temporal slopes) at the median and at $\tau = 0.99$ (Table~\ref{tab:2}). The correlation between random effects within weekdays or within weekends is strong, but the cross-correlation between weekdays and weekends terms is substantially weaker in the second half of the conditional distribution. This means that children tend to have trends of higher-intensity activity that are less similar between weekdays and weekends.

\begin{figure}[t!]
\centering
\includegraphics[scale = 0.4]{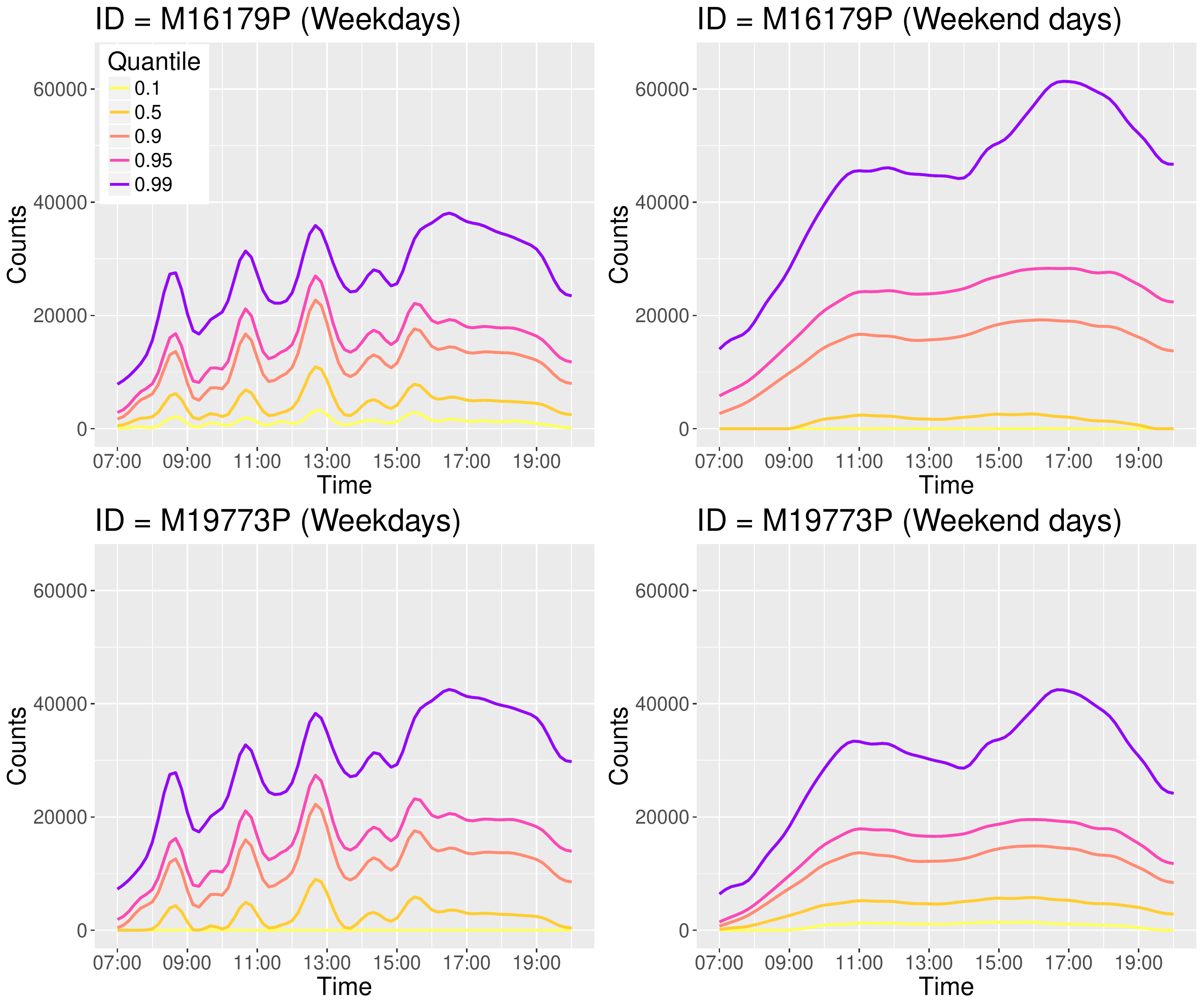}
\caption{Cluster-specific conditional quantile functions for two children (labelled M16179P and M19773P) of the UK Millennium Cohort Study physical activity dataset for 5 quantile levels $(0.1, 0.5, 0.9, 0.95, 0.99)$, by days of the week (Monday through Friday, weekdays; Saturday and Sunday, weekend).}
\label{fig:4}
\end{figure}

\textit{Individual} trajectories of accelerometer counts for two children of the MCS are plotted in Figure~\ref{fig:4}. Despite both being white females with similar BMI ($\sim 15.6$), living in a household with income in the lowest quintile and one car, having similar behaviors in terms of reading (often) and transportation (passive), they showed somewhat different daily patterns during summer weekend days. In particular, the conditional distribution was markedly skewed for the girl with identifier M16179P.

%\caption{Boxplots of the predicted random effects from the smooth terms for 5 quantile levels $(0.1, 0.5, 0.9, 0.95, 0.99)$ in the UK Millennium Cohort Study physical activity dataset.}

\section{Conclusion}
\label{sec:7}
We have developed a novel additive model for quantile regression when data are clustered. As compared to alternative approaches, ours has unique features, namely the mixed-effects representation of smoothing splines, which in turn leads to automatic smoothing selection, and the ability to model the variance-covariance matrix of the random effects.

As shown in a simulation study, the performance of AQMM was satisfactory despite the minimal tuning of the estimation algorithm. This takes a little burden away from the user who may instead focus their attention on other aspects of the analysis. This can be an asset if the data presents complexities as those illustrated in the MCS accelerometer analysis. In particular, the presence of a large number of regression coefficients and multiple smooth terms hinders the application of computationally intensive smoothing selection (e.g., cross-validation) to large datasets.

Standard error calculation in AQMM is facilitated by bootstrap. We were able to overcome the relatively large size of the MCS dataset by using an adaptation of the BLB approach \citep{kleiner_2014}. However, the versatility of bootstrap comes at a (computational) price and its application is limited to more central quantiles unless the sample size (i.e., number of clusters) is adequate. Further research is needed to develop accurate `sampling-free' approximations of standard errors in AQMM as well as in LQMM.

Finally, in contrast to estimation based on numerical quadrature \citep{geraci2014}, random effects estimates in AQMM are a by-product of the optimization algorithm rather than being calculated \textit{post hoc}. However, the proposed algorithm can be more demanding in terms of computing time as compared to, say, numerical quadrature or boosting, with the computational bottleneck indeed represented by the estimation of the random effects. For example, it took about two hours to fit a single AQMM using the MCS dataset. While, on the one hand, the large size of this dataset impaired even one of the most refined software for linear mixed-effects models, on the other hand a possible improvement in computing speed of the proposed algorithm is conceivable and is part of future research.

\section*{Acknowledgements}
This research has been supported by the National Institutes of Health -- National Institute of Child Health and Human Development (Grant Number: 1R03HD084807-01A1). The co-operation of the participating Millennium Cohort families is gratefully acknowledged as is the contribution of the management team at the Centre for Longitudinal Studies, UCL Institute of Education to collecting and, together with the UK Data Archive and Economic and Social Data Service, making the data available.

\clearpage

\appendix

\section*{Appendix A - Inference}
\renewcommand{\thesection}{A}
\numberwithin{equation}{section}
\setcounter{equation}{0}

The goal is to maximize the log-likelihood \eqref{eq:5} with respect to the $(p+m+1) \times 1$ parameter vector $\bm\theta_{\tau} = \left(\bm\beta_{\tau}\tp,\bm\xi_{\tau}\tp,\log{\bm\phi_{\tau}}\tp\right)\tp$.

Let $\mathbf{r}_{i} = \mathbf{y}_{i} - \mathbf{F}_{i}\bm\beta_{\tau} - \mathbf{Z}_{i}\mathbf{u}_{\tau,i} - \mathbf{B}_{i}\mathbf{v}_{\tau}$ be the $n_{i} \times 1$ vector of residuals for the $i$th cluster with generic element $r_{ij}$, and define the corresponding sign vector $\mathbf{s}_{i} = (s_{i1},\ldots,s_{in_{i}})\tp$ with
\begin{equation}\label{eq:A.1}
s_{ij} =
\begin{cases}
-1 & \text{if $r_{ij} \leq (\tau-1)\omega$},\\
0 & \text{if $(\tau - 1)\omega < r_{ij} < \tau
\omega$},\\
1 &  \text{if $r_{ij} \geq \tau\omega$}.
\end{cases}
\end{equation}
(The notation above has been simplified since the $r_{ij}$'s as well as the $s_{ij}$'s should be written as functions of $\bm\theta_{\tau}$.) We apply the smooth approximation \citep{madsen1993,chen2007} given in equation \eqref{eq:6} to the elements of $\mathbf{r}_{i}$ and write
\begin{equation}\label{eq:A.2}
\kappa_{\omega,\tau}(\mathbf{r}_{i}) \equiv \sum_{j=1}^{n_{i}} \kappa_{\omega,\tau}(r_{ij}) = \frac{1}{2}\left(\|\mathbf{r}_{i}\|^{2}_{\mathbf{A}_{i}} + \mathbf{b}_{i}\tp\mathbf{r}_{i} + \mathbf{c}_{i}\tp \mathbf{1}_{n_{i}}\right),
\end{equation}
where $\mathbf{A}_{i}$ is an $n_{i} \times n_{i}$ diagonal matrix with diagonal elements $\left\{\mathbf{A}_{i}\right\}_{jj} = (1 - s_{ij}^{2})/\omega$, $\mathbf{b}_{i}$ and $\mathbf{c}_{i}$ are two $n_{i} \times 1$ vectors with elements
\[
b_{ij} = s_{ij}((2\tau - 1)s_{ij} + 1)
\]
and
\[
c_{ij} = \frac{1}{2}\left\{(1 - 2\tau)\omega s_{ij} - (1 - 2\tau + 2\tau^2)\omega s_{ij}^2\right\},
\]
respectively.

For a more compact notation, let $\mathbf{w}_{\tau} =  \left(\mathbf{u}_{\tau}\tp, \mathbf{v}_{\tau}\tp\right)\tp$, $\mathbf{A} = \bigoplus_{i = 1}^{M} \mathbf{A}_{i}$, $\mathbf{b} =  \left(\mathbf{b}_{1}\tp, \ldots, \mathbf{b}_{M}\tp\right)\tp$, $\mathbf{c} =  \left(\mathbf{c}_{1}\tp, \ldots, \mathbf{c}_{M}\tp\right)\tp$, $\mathbf{r} =  \left(\mathbf{r}_{1}\tp, \ldots, \mathbf{r}_{M}\tp\right)\tp$, $\mathbf{Z} = \bigoplus_{i = 1}^{M} \mathbf{Z}_{i}$, $\mathbf{B} = \left[\begin{array}{ccc}\mathbf{B}_{1}\tp &\ldots & \mathbf{B}_{M}\tp \end{array}\right]\tp$, $\mathbf{G} = \left[\begin{array}{cc}\mathbf{Z} & \mathbf{B} \end{array}\right]$, and $\tilde{\bm\Psi}_{\tau} = (\mathbf{I}_{M} \otimes \tilde{\bm\Sigma})\,\oplus \,\tilde{\bm\Phi}$. We now define the function
\begin{align}\label{eq:A.3}
\nonumber h\left(\bm\theta_{\tau}, \mathbf{y}, \mathbf{u}_{\tau}, \mathbf{v}_{\tau}\right) &= \sum_{i=1}^{M}\left\{
\mathbf{r}_{i}\tp \mathbf{A}_{i} \mathbf{r}_{i} + \mathbf{b}_{i}\tp\mathbf{r}_{i} + \mathbf{c}_{i}\tp \mathbf{1}_{n_{i}}
+ \mathbf{u}_{\tau,i}\tp \tilde{\bm{\Sigma}}_{\tau}^{-1}\mathbf{u}_{\tau,i}\right\} + \mathbf{v}_{\tau}\tp \tilde{\bm\Phi}_{\tau}^{-1}\mathbf{v}_{\tau}\\
 & = \mathbf{r}\tp \mathbf{A} \mathbf{r} + \mathbf{b}\tp \mathbf{r} + \mathbf{c}\tp \mathbf{1}_{N} + \mathbf{w}_{\tau}\tp \tilde{\bm\Psi}_{\tau}^{-1}\mathbf{w}_{\tau}.
\end{align}
The smoothed version of the log-likelihood \eqref{eq:5} is then given by
\begin{align}\label{eq:A.4}
\nonumber \ell_{S}\left(\bm\theta_{\tau}; \mathbf{y}, \omega\right) &= N\log\left\{\frac{\tau(1-\tau)}{\sigma_{\tau}}\right\} - \frac{1}{2} \log |\tilde{\bm\Psi}_{\tau}|\\
& \quad + \log\int_{\mathbb{R}^{Mq+H}}\frac{\exp\left(-\frac{1}{2\sigma_{\tau}}h\left(\bm\theta_{\tau}, \mathbf{y}, \mathbf{u}_{\tau}, \mathbf{v}_{\tau}\right)\right)}{(2\pi\sigma_{\tau})^{(Mq+H)/2}} \rds \mathbf{w}_{\tau}.
\end{align}
For $\omega \rightarrow 0$, we have that $\ell_{S}\left(\bm\theta_{\tau}; \mathbf{y}, \omega\right) \rightarrow \ell\left(\bm\theta_{\tau}; \mathbf{y}\right)$.

Since $h$ is differentiable with respect to $\mathbf{w}_{\tau}$, we can derive the following quantities
\begin{align*}
\frac{\partial h\left(\bm\theta_{\tau}, \mathbf{y}, \mathbf{u}_{\tau}, \mathbf{v}_{\tau}\right)}{\partial \mathbf{u}_{\tau,i}} &= -\mathbf{Z}_{i}\tp \left(2\mathbf{A}_{i}\mathbf{r}_{i} + \mathbf{b}_{i}\right) + 2\tilde{\bm\Sigma}_{\tau}^{-1}\mathbf{u}_{\tau,i},\\
\frac{\partial h\left(\bm\theta_{\tau}, \mathbf{y}, \mathbf{u}_{\tau}, \mathbf{v}_{\tau}\right)}{\partial \mathbf{v}_{\tau}} &= -\sum_{i = 1}^{M}\left\{\mathbf{B}_{i}\tp \left(2\mathbf{A}_{i}\mathbf{r}_{i} + \mathbf{b}_{i}\right)\right\} + 2\tilde{\bm\Phi}_{\tau}^{-1}\mathbf{v}_{\tau},\\
\frac{\partial^{2} h\left(\bm\theta_{\tau}, \mathbf{y}, \mathbf{u}_{\tau}, \mathbf{v}_{\tau}\right)}{\partial \mathbf{u}_{\tau,i}\mathbf{u}_{\tau,i}\tp} &= 2\left(\mathbf{Z}_{i}\tp\mathbf{A}_{i}\mathbf{Z}_{i} + \tilde{\bm\Sigma}_{\tau}^{-1}\right),\\
\frac{\partial^{2} h\left(\bm\theta_{\tau}, \mathbf{y}, \mathbf{u}_{\tau}, \mathbf{v}_{\tau}\right)}{\partial \mathbf{v}_{\tau}\mathbf{v}_{\tau}\tp} &= \sum_{i = 1}^{M} 2\left(\mathbf{B}_{i}\tp\mathbf{A}_{i}\mathbf{B}_{i}\right) + 2\tilde{\bm\Phi}_{\tau}^{-1},\\
\frac{\partial^{2} h\left(\bm\theta_{\tau}, \mathbf{y}, \mathbf{u}_{\tau}, \mathbf{v}_{\tau}\right)}{\partial \mathbf{u}_{\tau,i}\mathbf{u}_{\tau,j}\tp} &= \mathbf{O}_{q \times q}, \qquad i \neq j,\\
\frac{\partial^{2} h\left(\bm\theta_{\tau}, \mathbf{y}, \mathbf{u}_{\tau}, \mathbf{v}_{\tau}\right)}{\partial \mathbf{v}_{\tau}\mathbf{u}_{\tau,i}\tp} &= 2\mathbf{B}_{i}\tp\mathbf{A}_{i}\mathbf{Z}_{i}. \\
\end{align*}
The above derivatives can be written more compactly for all clusters as
\begin{align}
\label{eq:A.5} \frac{\partial h\left(\bm\theta_{\tau}, \mathbf{y}, \mathbf{u}_{\tau}, \mathbf{v}_{\tau}\right)}{\partial \mathbf{w}_{\tau}} &= -\mathbf{G}\tp\left(2\mathbf{A}\mathbf{r} + \mathbf{b}\right) + 2\tilde{\bm\Psi}^{-1}_{\tau}\mathbf{w}_{\tau},\\
\label{eq:A.6} \frac{\partial^{2} h\left(\bm\theta_{\tau}, \mathbf{y}, \mathbf{u}_{\tau}, \mathbf{v}_{\tau}\right)}{\partial \mathbf{w}_{\tau}\mathbf{w}_{\tau}\tp} &=2\left(\mathbf{G}\tp \mathbf{A} \mathbf{G} + \tilde{\bm\Psi}^{-1}_{\tau}\right).
\end{align}

Moreover, let
\begin{equation}\label{eq:A.7}
\hat{\mathbf{w}}_{\tau} \equiv (\hat{\mathbf{u}}_{\tau}, \hat{\mathbf{v}}_{\tau}) = \argmin_{\mathbf{u},\mathbf{v}} h\left(\bm\theta_{\tau}, \mathbf{y}, \mathbf{u}, \mathbf{v}\right)
\end{equation}
be the conditional mode of $\mathbf{w}_{\tau}$. A second-order approximation of $h$ around $\hat{\mathbf{w}}_{\tau}$ is given by
\begin{equation*}
h\left(\bm\theta_{\tau}, \mathbf{y}, \mathbf{u}_{\tau}, \mathbf{v}_{\tau}\right) \simeq
h_{0} +
\dot{\mathbf{h}}\tp \left(\mathbf{w}_{\tau}-\hat{\mathbf{w}}_{\tau}\right) +
\left(\mathbf{w}_{\tau}-\hat{\mathbf{w}}_{\tau}\right)\tp \ddot{\mathbf{H}} \left(\mathbf{w}_{\tau}-\hat{\mathbf{w}}_{\tau}\right),
\end{equation*}
where $h_{0}\equiv h\left(\bm\theta_{\tau}, \mathbf{y}, \hat{\mathbf{u}}_{\tau}, \hat{\mathbf{v}}_{\tau}\right)$, $\dot{\mathbf{h}}\equiv
h'\left(\bm\theta_{\tau}, \mathbf{y}, \hat{\mathbf{u}}_{\tau}, \hat{\mathbf{v}}_{\tau}\right)$, and $\ddot{\mathbf{H}}\equiv h''\left(\bm\theta_{\tau}, \mathbf{y}, \hat{\mathbf{u}}_{\tau}, \hat{\mathbf{v}}_{\tau}\right)/2$. Since $\dot{\mathbf{h}}$ is zero at $\mathbf{w}_{\tau} = \hat{\mathbf{w}}_{\tau}$, we have finally the following Laplacian approximation of the (smoothed) log-likelihood \eqref{eq:A.4}
\begin{align}\label{eq:A.8}
\nonumber \ell_{\mathrm{LA}}\left(\bm\theta_{\tau}; \mathbf{y}, \hat{\mathbf{w}}_{\tau}\right) & = N\log\left\{\frac{\tau(1-\tau)}{\sigma_{\tau}}\right\} - \frac{1}{2} \log |\tilde{\bm\Psi}_{\tau}| -\frac{1}{2\sigma_{\tau}}h_{0} \\
\nonumber &\quad + \log \int_{\mathbb{R}^{Mq+H}} (2\pi \sigma_{\tau})^{-(Mq+H)/2} \exp\left\{-\frac{1}{2\sigma_{\tau}} \left(\mathbf{w}_{\tau}-\hat{\mathbf{w}}_{\tau}\right)\tp \ddot{\mathbf{H}} \left(\mathbf{w}_{\tau}-\hat{\mathbf{w}}_{\tau}\right) \right\}\rds \mathbf{w}_{\tau}\\
& = N\log\left\{\frac{\tau (1 - \tau)}{\sigma_{\tau}}\right\} -\frac{1}{2} \left(\log |\tilde{\bm\Psi}_{\tau}\ddot{\mathbf{H}}| + \sigma_{\tau}^{-1}h_{0}\right).
\end{align}

To maximize \eqref{eq:A.8} with respect to $\bm\theta_{\tau}$, we can use a general purpose optimizer such the Nelder--Mead or the Broyden--Fletcher--Goldfarb--Shanno algorithms. For a given $\hat{\bm\theta}_{\tau}$, the scale $\sigma_{\tau}$ can be estimated from
\begin{equation}\label{eq:A.9}
\hat{\sigma}_{\tau} = (2N)^{-1} h_{0}.
\end{equation}
Note that \eqref{eq:A.8} can be profiled with respect to $\sigma_{\tau}$. Finally, for a given value of $\omega$, the mode $\mathbf{w}_{\tau}$ can be obtained by means of Newton-Raphson \citep{pinheiro_chao} using \eqref{eq:A.5} and \eqref{eq:A.6}.

Estimation of the parameters can be carried out iteratively. The algorithm requires setting the starting value of $\bm\theta_{\tau}$ and $\sigma_{\tau}$, the tuning parameter $\omega$, the tolerance for the change in the log-likelihood, and the maximum number of iterations. Moreover, the modes of the random effects can be obtained by equating \eqref{eq:A.5} to $\mathbf{0}$ and then solving for $\mathbf{w}_{\tau}$. Specifically, this leads to the following system of equations
\begin{equation}\label{eq:A.10}
2\ddot{\mathbf{H}}\mathbf{w}_{\tau} = \mathbf{G}\tp\left\{2\mathbf{A}(\mathbf{y} - \mathbf{F}\bm\beta_{\tau})+ \mathbf{b}\right\}.
\end{equation}
Since the right-hand side depends on $\mathbf{w}_{\tau}$ through $\mathbf{A}$ and $\mathbf{b}$, an estimate $\hat{\mathbf{w}}_{\tau}$ is obtained iteratively.

The pseudo-code of the complete algorithm is given below. This algorithm is based on a set of decreasing values of $\omega$ and has the appealing advantage of reducing the original non-smooth problem to an approximated $L_{2}$ problem.

\clearpage
\begin{framed}
\begin{center}
\textsc{Smoothing Algorithm with Laplacian Approximation for Additive Quantile Mixed Models}
\begin{itemize}
\item[(1)] Set the maximum number of iterations $T$; the factor $0< \gamma < 1$ for reducing the tuning parameter $\omega$; the tolerance for the change in the log-likelihood; and $t = 0$. Estimate the starting values as follows:
\begin{itemize}
\item[(a)] obtain an estimate for $\bm\beta^{(0)}_{\tau}$ using an additive mixed model (AMM) \citep{wood_2006b}. If the AMM fitting algorithm fails, consider a standard linear least squares estimate of the fixed effects;
\item[(b)] obtain an estimate for $\bm\xi^{(0)}_{\tau}$ and $\bm\phi^{(0)}_{\tau}$ from the AMM fitted in step (a). If the AMM fitting algorithm fails, provide an arbitrary value;
\item[(c)] obtain an estimate for $\sigma_{\tau}^{(0)}$. For example, this can be estimated as the mean of the absolute residuals from step (1.a) above;
\item[(d)] provide a starting value $\omega^{(0)}$ \citep[see, for example,][p.143]{chen2007};
\item[(e)] using $\bm\beta^{(0)}_{\tau}$, $\bm\xi^{(0)}_{\tau}$, $\bm\phi^{(0)}_{\tau}$, and $\sigma^{(0)}_{\tau}$, solve \eqref{eq:A.7} to obtain $\mathbf{w}_{\tau}^{(0)}$ using an iterative method. See, for example, the \texttt{R} functions \texttt{optim} and \texttt{nlm}.
\end{itemize}
\item[(2)] While $t < T$
\begin{itemize}
\item[(a)] Update $\bm\theta_{\tau}^{(t)}$ by minimizing \eqref{eq:A.8}. See, for example, the \texttt{R} function \texttt{optim}.
\item[(b)] If the change in the log-likelihood is smaller than a given tolerance
\begin{itemize}
    \item[(i)] then return $\bm\theta_{\tau}^{(t + 1)}$;
    \item[(ii)] else set $\bm\theta_{\tau}^{(t + 1)} = \bm\theta_{\tau}^{(t)}$; $\omega^{(t+1)} = \gamma \cdot \omega^{(t)}$; $t = t + 1$; go to step (2.a).
    \end{itemize}
\end{itemize}
\item[(3)] Update $\sigma^{(t)}_{\tau}$ and $\mathbf{w}^{(t)}_{\tau}$.
\end{itemize}

\end{center}
\end{framed}

\clearpage

\section*{Appendix B - Simulation study results}
\renewcommand{\thesection}{B}
\numberwithin{equation}{section}
\setcounter{equation}{0}
\renewcommand{\thetable}{B\arabic{table}}
\setcounter{table}{0}

\subsection{Homoscedastic scenario}

\begin{table}[ht]
\caption{Quantile 0.1 and homoscedastic scenario. Average bias and root mean squared error (RMSE) for $Q(0.1)$ from the additive quantile mixed model (AQMM) and the additive fixed-effects quantile regression (AFEQR). The expected proportion of negative residuals (PNR) is 0.1.}
\centering
\begin{tabular}{lrrrrrr}
  \toprule
Sample size & \multicolumn{3}{c}{AQMM} & \multicolumn{3}{c}{AFEQR} \\
($n$, $M$) & PNR & Bias & RMSE & PNR & Bias & RMSE \\
  \midrule
  \multicolumn{7}{l}{\textit{Normal}}\\
  \midrule
(5, 50) & 0.11 & 0.27 & 1.69 & 0.10 & $-$2.21 & 3.22 \\
  (10, 50) & 0.10 & 0.12 & 1.71 & 0.10 & $-$2.34 & 3.26 \\
  (5, 100) & 0.10 & 0.26 & 1.69 & 0.10 & $-$2.35 & 3.28 \\
  (10, 100) & 0.10 & 0.11 & 1.69 & 0.10 & $-$2.40 & 3.24 \\
  (5, 500) & 0.10 & 0.23 & 1.67 & 0.10 & $-$2.36 & 3.13 \\
  (10, 500) & 0.10 & 0.13 & 1.69 & 0.10 & $-$2.36 & 3.12 \\
  \midrule
  \multicolumn{7}{l}{\textit{Student's $t$}}\\
  \midrule
  (5, 50) & 0.10 & 0.01 & 1.65 & 0.10 & $-$2.14 & 3.25 \\
  (10, 50) & 0.10 & $-$0.02 & 1.66 & 0.10 & $-$2.25 & 3.25 \\
  (5, 100) & 0.10 & 0.06 & 1.60 & 0.10 & $-$2.29 & 3.29 \\
  (10, 100) & 0.10 & 0.02 & 1.64 & 0.10 & $-$2.32 & 3.23 \\
  (5, 500) & 0.10 & 0.06 & 1.54 & 0.10 & $-$2.29 & 3.10 \\
  (10, 500) & 0.10 & 0.02 & 1.61 & 0.10 & $-$2.29 & 3.10 \\
  \midrule
  \multicolumn{7}{l}{\textit{Chi-squared}}\\
  \midrule
  (5, 50) & 0.10 & $-$0.01 & 1.56 & 0.10 & $-$1.81 & 3.02 \\
  (10, 50) & 0.09 & $-$0.02 & 1.55 & 0.10 & $-$1.94 & 3.01 \\
  (5, 100) & 0.10 & 0.01 & 1.47 & 0.10 & $-$1.98 & 3.06 \\
  (10, 100) & 0.09 & $-$0.01 & 1.54 & 0.10 & $-$2.03 & 2.99 \\
  (5, 500) & 0.10 & 0.04 & 1.40 & 0.10 & $-$2.01 & 2.88 \\
  (10, 500) & 0.10 & 0.03 & 1.51 & 0.10 & $-$2.01 & 2.88 \\
  \bottomrule
\end{tabular}
\label{tab:B.1}
\end{table}

\begin{table}[ht]
\caption{Quantile 0.5 and homoscedastic scenario. Average bias and root mean squared error (RMSE) for $Q(0.5)$ from the additive quantile mixed model (AQMM), the additive fixed-effects quantile regression (AFEQR), and the additive mixed model (AMM). The expected proportion of negative residuals (PNR) is 0.5.}
\centering
\begin{tabular}{lrrrrrrrrr}
  \toprule
Sample size & \multicolumn{3}{c}{AQMM} & \multicolumn{3}{c}{AFEQR} & \multicolumn{3}{c}{AMM}\\
($n$, $M$) & PNR & Bias & RMSE & PNR & Bias & RMSE & PNR & Bias & RMSE\\
  \midrule
  \multicolumn{10}{l}{\textit{Normal}}\\
  \midrule
(5, 50) & 0.50 & $-$0.00 & 1.66 & 0.50 & $-$0.01 & 1.69 & 0.50 & $-$0.00 & 1.65 \\
  (10, 50) & 0.50 & 0.00 & 1.69 & 0.50 & $-$0.02 & 1.50 & 0.50 & $-$0.00 & 1.69 \\
  (5, 100) & 0.50 & 0.01 & 1.66 & 0.50 & 0.00 & 1.50 & 0.50 & 0.01 & 1.66 \\
  (10, 100) & 0.50 & $-$0.00 & 1.68 & 0.50 & $-$0.01 & 1.36 & 0.50 & 0.00 & 1.69 \\
  (5, 500) & 0.50 & 0.00 & 1.65 & 0.50 & $-$0.02 & 1.24 & 0.50 & 0.00 & 1.65 \\
  (10, 500) & 0.50 & 0.00 & 1.69 & 0.50 & $-$0.01 & 1.24 & 0.50 & 0.00 & 1.68 \\
  \midrule
  \multicolumn{10}{l}{\textit{Student's $t$}}\\
  \midrule
  (5, 50) & 0.50 & $-$0.00 & 1.58 & 0.50 & $-$0.00 & 1.79 & 0.50 & $-$0.01 & 1.60 \\
  (10, 50) & 0.50 & $-$0.01 & 1.61 & 0.50 & $-$0.03 & 1.57 & 0.50 & $-$0.01 & 1.62 \\
  (5, 100) & 0.50 & 0.00 & 1.56 & 0.50 & $-$0.01 & 1.58 & 0.50 & $-$0.00 & 1.57 \\
  (10, 100) & 0.50 & 0.00 & 1.61 & 0.50 & $-$0.01 & 1.43 & 0.50 & 0.00 & 1.62 \\
  (5, 500) & 0.50 & 0.00 & 1.53 & 0.50 & $-$0.01 & 1.30 & 0.50 & 0.00 & 1.53 \\
  (10, 500) & 0.50 & 0.00 & 1.61 & 0.50 & $-$0.02 & 1.30 & 0.50 & 0.00 & 1.61 \\
  \midrule
  \multicolumn{10}{l}{\textit{Chi-squared}}\\
  \midrule
 (5, 50) & 0.50 & 0.31 & 1.59 & 0.50 & 0.40 & 2.05 & 0.88 & 2.99 & 3.37 \\
  (10, 50) & 0.50 & 0.24 & 1.59 & 0.50 & 0.37 & 1.80 & 0.87 & 3.01 & 3.39 \\
  (5, 100) & 0.50 & 0.28 & 1.51 & 0.50 & 0.37 & 1.83 & 0.87 & 3.00 & 3.34 \\
  (10, 100) & 0.50 & 0.19 & 1.55 & 0.50 & 0.33 & 1.64 & 0.86 & 3.00 & 3.37 \\
  (5, 500) & 0.50 & 0.25 & 1.43 & 0.50 & 0.32 & 1.48 & 0.87 & 3.00 & 3.31 \\
  (10, 500) & 0.50 & 0.18 & 1.52 & 0.50 & 0.33 & 1.47 & 0.86 & 3.00 & 3.36 \\
   \bottomrule
\end{tabular}
\label{tab:B.2}
\end{table}

\begin{table}[ht]
\caption{Quantile 0.95 and homoscedastic scenario. Average bias and root mean squared error (RMSE) for $Q(0.95)$ from the additive quantile mixed model (AQMM) and the additive fixed-effects quantile regression (AFEQR). The expected proportion of negative residuals (PNR) is 0.95.}
\centering
\begin{tabular}{lrrrrrr}
  \toprule
Sample size & \multicolumn{3}{c}{AQMM} & \multicolumn{3}{c}{AFEQR} \\
($n$, $M$) & PNR & Bias & RMSE & PNR & Bias & RMSE \\
  \midrule
  \multicolumn{7}{l}{\textit{Normal}}\\
  \midrule
(5, 50) & 0.94 & $-$0.33 & 1.72 & 0.95 & 3.03 & 4.01 \\
  (10, 50) & 0.95 & $-$0.16 & 1.72 & 0.95 & 3.22 & 4.12 \\
  (5, 100) & 0.95 & $-$0.29 & 1.70 & 0.95 & 3.27 & 4.17 \\
  (10, 100) & 0.95 & $-$0.15 & 1.70 & 0.95 & 3.32 & 4.17 \\
  (5, 500) & 0.95 & $-$0.28 & 1.68 & 0.95 & 3.30 & 4.08 \\
  (10, 500) & 0.95 & $-$0.15 & 1.70 & 0.95 & 3.30 & 4.08 \\
  \midrule
  \multicolumn{7}{l}{\textit{Student's $t$}}\\
  \midrule
  (5, 50) & 0.94 & $-$0.16 & 1.76 & 0.95 & 2.74 & 3.86 \\
  (10, 50) & 0.95 & $-$0.09 & 1.73 & 0.95 & 2.90 & 3.92 \\
  (5, 100) & 0.95 & $-$0.19 & 1.68 & 0.95 & 2.96 & 3.97 \\
  (10, 100) & 0.95 & $-$0.11 & 1.68 & 0.95 & 3.02 & 3.96 \\
  (5, 500) & 0.95 & $-$0.19 & 1.57 & 0.95 & 2.99 & 3.85 \\
  (10, 500) & 0.95 & $-$0.11 & 1.63 & 0.95 & 2.99 & 3.85 \\
  \midrule
  \multicolumn{7}{l}{\textit{Chi-squared}}\\
  \midrule
  (5, 50) & 0.95 & $-$0.70 & 2.18 & 0.95 & 1.49 & 3.36 \\
  (10, 50) & 0.95 & $-$0.43 & 2.01 & 0.95 & 1.79 & 3.37 \\
  (5, 100) & 0.95 & $-$0.68 & 1.97 & 0.95 & 1.82 & 3.39 \\
  (10, 100) & 0.95 & $-$0.44 & 1.86 & 0.95 & 1.91 & 3.33 \\
  (5, 500) & 0.95 & $-$0.65 & 1.67 & 0.95 & 1.86 & 3.15 \\
  (10, 500) & 0.95 & $-$0.43 & 1.65 & 0.95 & 1.86 & 3.12 \\
  \bottomrule
\end{tabular}
\label{tab:B.3}
\end{table}

\begin{table}[ht]
\caption{Quantile 0.1 and homoscedastic scenario. Average relative bias (\%) and root mean squared error (RMSE) for $\beta_3$ from the additive quantile mixed model (AQMM) and the additive fixed-effects quantile regression (AFEQR).}
\centering
\begin{tabular}{lrrrr}
  \toprule
Sample size & \multicolumn{2}{c}{AQMM} & \multicolumn{2}{c}{AFEQR} \\
($n$, $M$) & Rel. bias & RMSE & Rel. bias & RMSE \\
  \midrule
  \multicolumn{5}{l}{\textit{Normal}}\\
  \midrule
(5, 50) & $-$0.17 & 0.05 & $-$6.05 & 0.43 \\
  (10, 50) & $-$0.22 & 0.03 & $-$8.42 & 0.29 \\
  (5, 100) & 0.02 & 0.03 & $-$8.25 & 0.29 \\
  (10, 100) & 0.20 & 0.01 & $-$8.91 & 0.22 \\
  (5, 500) & $-$0.15 & 0.01 & $-$9.23 & 0.17 \\
  (10, 500) & 0.05 & 0.00 & $-$9.14 & 0.15 \\
  \midrule
  \multicolumn{5}{l}{\textit{Student's $t$}}\\
  \midrule
  (5, 50) & $-$0.54 & 0.17 & $-$7.17 & 0.52 \\
  (10, 50) & 0.01 & 0.07 & $-$8.92 & 0.35 \\
  (5, 100) & $-$0.40 & 0.07 & $-$9.23 & 0.38 \\
  (10, 100) & 0.09 & 0.04 & $-$8.98 & 0.24 \\
  (5, 500) & 0.09 & 0.01 & $-$9.51 & 0.19 \\
  (10, 500) & $-$0.06 & 0.01 & $-$10.12 & 0.19 \\
  \midrule
  \multicolumn{5}{l}{\textit{Chi-squared}}\\
  \midrule
  (5, 50) & $-$0.64 & 0.16 & $-$7.81 & 0.50 \\
  (10, 50) & $-$0.21 & 0.06 & $-$8.87 & 0.35 \\
  (5, 100) & $-$0.21 & 0.07 & $-$9.49 & 0.36 \\
  (10, 100) & $-$0.01 & 0.03 & $-$9.97 & 0.27 \\
  (5, 500) & 0.20 & 0.01 & $-$8.97 & 0.17 \\
  (10, 500) & $-$0.00 & 0.00 & $-$9.94 & 0.18 \\
   \bottomrule
 \end{tabular}
\label{tab:B.4}
\end{table}

\begin{table}[ht]
\caption{Quantile 0.5 and homoscedastic scenario. Average relative bias (\%) and root mean squared error (RMSE) for $\beta_3$ from the additive quantile mixed model (AQMM), the additive fixed-effects quantile regression (AFEQR), and the additive mixed model (AMM).}
\centering
\begin{tabular}{lrrrrrr}
  \toprule
Sample size & \multicolumn{2}{c}{AQMM} & \multicolumn{2}{c}{AFEQR} & \multicolumn{2}{c}{AMM}\\
($n$, $M$) & Rel. bias & RMSE & Rel. bias & RMSE & Rel. bias & RMSE\\
  \midrule
  \multicolumn{7}{l}{\textit{Normal}}\\
  \midrule
(5, 50) & $-$0.50 & 0.04 & $-$4.81 & 0.26 & $-$0.53 & 0.13 \\
  (10, 50) & $-$0.13 & 0.02 & $-$5.36 & 0.15 & $-$0.10 & 0.09 \\
  (5, 100) & 0.17 & 0.02 & $-$4.70 & 0.14 & 0.16 & 0.10 \\
  (10, 100) & 0.03 & 0.01 & $-$4.70 & 0.08 & 0.02 & 0.06 \\
  (5, 500) & $-$0.11 & 0.00 & $-$4.40 & 0.05 & $-$0.11 & 0.04 \\
  (10, 500) & 0.04 & 0.00 & $-$4.59 & 0.04 & 0.00 & 0.03 \\
  \midrule
  \multicolumn{7}{l}{\textit{Student's $t$}}\\
  \midrule
  (5, 50) & $-$0.37 & 0.08 & $-$4.09 & 0.26 & $-$0.20 & 0.22 \\
  (10, 50) & $-$0.00 & 0.03 & $-$5.88 & 0.16 & $-$0.25 & 0.14 \\
  (5, 100) & $-$0.20 & 0.03 & $-$4.86 & 0.15 & $-$0.51 & 0.15 \\
  (10, 100) & $-$0.05 & 0.01 & $-$5.11 & 0.09 & 0.17 & 0.10 \\
  (5, 500) & $-$0.05 & 0.01 & $-$4.57 & 0.05 & $-$0.06 & 0.06 \\
  (10, 500) & $-$0.02 & 0.00 & $-$4.98 & 0.05 & $-$0.07 & 0.04 \\
  \midrule
  \multicolumn{7}{l}{\textit{Chi-squared}}\\
  \midrule
  (5, 50) & 0.07 & 0.15 & $-$4.05 & 0.34 & $-$0.07 & 0.30 \\
  (10, 50) & $-$0.24 & 0.08 & $-$5.60 & 0.21 & $-$0.45 & 0.21 \\
  (5, 100) & 0.16 & 0.08 & $-$5.22 & 0.18 & $-$0.05 & 0.23 \\
  (10, 100) & $-$0.25 & 0.04 & $-$5.45 & 0.13 & $-$0.31 & 0.15 \\
  (5, 500) & 0.23 & 0.02 & $-$4.56 & 0.06 & 0.16 & 0.09 \\
  (10, 500) & $-$0.02 & 0.01 & $-$4.65 & 0.05 & $-$0.06 & 0.06 \\
   \bottomrule
\end{tabular}
\label{tab:B.5}
\end{table}

\begin{table}[ht]
\caption{Quantile 0.95 and homoscedastic scenario. Average relative bias (\%) and root mean squared error (RMSE) for $\beta_3$ from the additive quantile mixed model (AQMM) and the additive fixed-effects quantile regression (AFEQR).}
\centering
\begin{tabular}{lrrrr}
  \toprule
Sample size & \multicolumn{2}{c}{AQMM} & \multicolumn{2}{c}{AFEQR} \\
($n$, $M$) & Rel. bias & RMSE & Rel. bias & RMSE \\
  \midrule
  \multicolumn{5}{l}{\textit{Normal}}\\
  \midrule
(5, 50) & $-$1.15 & 0.07 & $-$7.67 & 0.59 \\
  (10, 50) & $-$0.38 & 0.04 & $-$7.08 & 0.33 \\
  (5, 100) & 0.45 & 0.04 & $-$7.66 & 0.36 \\
  (10, 100) & $-$0.07 & 0.02 & $-$6.14 & 0.18 \\
  (5, 500) & $-$0.13 & 0.01 & $-$5.54 & 0.10 \\
  (10, 500) & $-$0.07 & 0.00 & $-$5.45 & 0.07 \\
  \midrule
  \multicolumn{5}{l}{\textit{Student's $t$}}\\
  \midrule
  (5, 50) & 0.86 & 0.33 & $-$6.43 & 0.74 \\
  (10, 50) & $-$0.44 & 0.16 & $-$7.31 & 0.37 \\
  (5, 100) & $-$0.03 & 0.13 & $-$7.66 & 0.41 \\
  (10, 100) & 0.54 & 0.08 & $-$7.43 & 0.26 \\
  (5, 500) & $-$0.15 & 0.03 & $-$6.35 & 0.13 \\
  (10, 500) & $-$0.41 & 0.02 & $-$6.21 & 0.09 \\
  \midrule
  \multicolumn{5}{l}{\textit{Chi-squared}}\\
  \midrule
  (5, 50) & $-$0.57 & 1.17 & $-$14.98 & 1.55 \\
  (10, 50) & $-$0.75 & 0.67 & $-$13.73 & 0.95 \\
  (5, 100) & 0.50 & 0.70 & $-$12.12 & 0.94 \\
  (10, 100) & 0.39 & 0.37 & $-$11.47 & 0.56 \\
  (5, 500) & 0.29 & 0.14 & $-$9.29 & 0.29 \\
  (10, 500) & $-$0.44 & 0.07 & $-$8.98 & 0.20 \\
   \bottomrule
 \end{tabular}
\label{tab:B.6}
\end{table}

\begin{table}[ht]
\caption{Quantile 0.1 and homoscedastic scenario. Average relative bias (\%) and root mean squared error (RMSE) for $\beta_4$ from the additive quantile mixed model (AQMM) and the additive fixed-effects quantile regression (AFEQR).}
\centering
\begin{tabular}{lrrrr}
  \toprule
Sample size & \multicolumn{2}{c}{AQMM} & \multicolumn{2}{c}{AFEQR} \\
($n$, $M$) & Rel. bias & RMSE & Rel. bias & RMSE \\
  \midrule
  \multicolumn{5}{l}{\textit{Normal}}\\
  \midrule
(5, 50) & 0.59 & 0.03 & $-$15.62 & 0.31 \\
  (10, 50) & 0.40 & 0.03 & $-$15.49 & 0.28 \\
  (5, 100) & 0.60 & 0.02 & $-$15.55 & 0.27 \\
  (10, 100) & 0.37 & 0.02 & $-$15.45 & 0.25 \\
  (5, 500) & 0.85 & 0.00 & $-$15.18 & 0.22 \\
  (10, 500) & 0.28 & 0.00 & $-$15.40 & 0.22 \\
  \midrule
  \multicolumn{5}{l}{\textit{Student's $t$}}\\
  \midrule
  (5, 50) & 1.70 & 0.06 & $-$16.18 & 0.37 \\
  (10, 50) & 0.60 & 0.03 & $-$15.61 & 0.28 \\
  (5, 100) & 1.38 & 0.03 & $-$15.97 & 0.29 \\
  (10, 100) & 0.04 & 0.02 & $-$15.50 & 0.25 \\
  (5, 500) & 1.52 & 0.01 & $-$15.35 & 0.22 \\
  (10, 500) & 0.55 & 0.00 & $-$15.49 & 0.22 \\
  \midrule
  \multicolumn{5}{l}{\textit{Chi-squared}}\\
  \midrule
  (5, 50) & 2.13 & 0.08 & $-$15.03 & 0.34 \\
  (10, 50) & 0.53 & 0.03 & $-$14.82 & 0.27 \\
  (5, 100) & 1.63 & 0.03 & $-$15.03 & 0.27 \\
  (10, 100) & 0.43 & 0.02 & $-$15.36 & 0.25 \\
  (5, 500) & 1.27 & 0.01 & $-$14.81 & 0.21 \\
  (10, 500) & 0.40 & 0.00 & $-$15.18 & 0.21 \\
   \bottomrule
\end{tabular}
\label{tab:B.7}
\end{table}

\begin{table}[ht]
\caption{Quantile 0.5 and homoscedastic scenario. Average relative bias (\%) and root mean squared error (RMSE) for $\beta_4$ from the additive quantile mixed model (AQMM), the additive fixed-effects quantile regression (AFEQR), and the additive mixed model (AMM).}
\centering
\begin{tabular}{lrrrrrr}
  \toprule
 Sample size  & \multicolumn{2}{c}{AQMM} & \multicolumn{2}{c}{AFEQR} & \multicolumn{2}{c}{AMM}\\
   ($n$, $M$) & Rel. bias & RMSE & Rel. bias & RMSE & Rel. bias & RMSE\\
  \midrule
  \multicolumn{7}{l}{\textit{Normal}}\\
  \midrule
(5, 50) & $-$0.26 & 0.03 & $-$5.64 & 0.09 & $-$0.24 & 0.13 \\
  (10, 50) & $-$0.24 & 0.02 & $-$6.61 & 0.08 & $-$0.32 & 0.12 \\
  (5, 100) & $-$0.09 & 0.01 & $-$5.29 & 0.05 & $-$0.17 & 0.09 \\
  (10, 100) & $-$0.13 & 0.01 & $-$5.69 & 0.05 & $-$0.04 & 0.09 \\
  (5, 500) & 0.11 & 0.00 & $-$5.26 & 0.03 & 0.12 & 0.04 \\
  (10, 500) & 0.01 & 0.00 & $-$5.51 & 0.03 & 0.01 & 0.04 \\
  \midrule
  \multicolumn{7}{l}{\textit{Student's $t$}}\\
  \midrule
  (5, 50) & $-$0.12 & 0.04 & $-$5.59 & 0.10 & $-$0.07 & 0.16 \\
  (10, 50) & $-$0.09 & 0.02 & $-$6.50 & 0.08 & $-$0.11 & 0.12 \\
  (5, 100) & $-$0.02 & 0.02 & $-$5.30 & 0.06 & 0.08 & 0.11 \\
  (10, 100) & $-$0.15 & 0.01 & $-$6.23 & 0.06 & $-$0.18 & 0.09 \\
  (5, 500) & 0.09 & 0.00 & $-$5.48 & 0.03 & 0.10 & 0.05 \\
  (10, 500) & 0.08 & 0.00 & $-$5.59 & 0.03 & 0.09 & 0.04 \\
  \midrule
  \multicolumn{7}{l}{\textit{Chi-squared}}\\
  \midrule
  (5, 50) & 1.72 & 0.07 & $-$5.79 & 0.12 & $-$0.04 & 0.20 \\
  (10, 50) & 0.67 & 0.04 & $-$6.85 & 0.09 & $-$0.27 & 0.15 \\
  (5, 100) & 1.71 & 0.03 & $-$5.43 & 0.07 & 0.21 & 0.13 \\
  (10, 100) & 0.59 & 0.02 & $-$5.75 & 0.06 & $-$0.35 & 0.11 \\
  (5, 500) & 1.64 & 0.01 & $-$4.97 & 0.03 & 0.02 & 0.05 \\
  (10, 500) & 1.05 & 0.00 & $-$5.22 & 0.03 & 0.04 & 0.05 \\
   \bottomrule
\end{tabular}
\label{tab:B.8}
\end{table}

\begin{table}[ht]
\caption{Quantile 0.95 and homoscedastic scenario. Average relative bias (\%) and root mean squared error (RMSE) for $\beta_4$ from the additive quantile mixed model (AQMM) and the additive fixed-effects quantile regression (AFEQR).}
\centering
\begin{tabular}{lrrrr}
  \toprule
Sample size & \multicolumn{2}{c}{AQMM} & \multicolumn{2}{c}{AFEQR} \\
($n$, $M$) & Rel. bias & RMSE & Rel. bias & RMSE \\
  \midrule
  \multicolumn{5}{l}{\textit{Normal}}\\
  \midrule
(5, 50) & $-$0.98 & 0.04 & 0.25 & 0.16 \\
  (10, 50) & $-$0.96 & 0.04 & 2.39 & 0.09 \\
  (5, 100) & $-$1.18 & 0.02 & 4.13 & 0.09 \\
  (10, 100) & $-$0.68 & 0.02 & 4.18 & 0.06 \\
  (5, 500) & $-$0.78 & 0.00 & 5.31 & 0.04 \\
  (10, 500) & $-$0.31 & 0.00 & 5.21 & 0.03 \\
  \midrule
  \multicolumn{5}{l}{\textit{Student's $t$}}\\
  \midrule
  (5, 50) & $-$2.29 & 0.08 & $-$0.26 & 0.22 \\
  (10, 50) & $-$1.18 & 0.05 & $-$0.00 & 0.11 \\
  (5, 100) & $-$1.46 & 0.04 & 1.94 & 0.10 \\
  (10, 100) & $-$0.98 & 0.03 & 1.83 & 0.06 \\
  (5, 500) & $-$1.60 & 0.01 & 4.35 & 0.03 \\
  (10, 500) & $-$0.63 & 0.01 & 4.11 & 0.03 \\
  \midrule
  \multicolumn{5}{l}{\textit{Chi-squared}}\\
  \midrule
  (5, 50) & $-$6.70 & 0.30 & $-$10.24 & 0.49 \\
  (10, 50) & $-$3.23 & 0.17 & $-$6.72 & 0.25 \\
  (5, 100) & $-$5.68 & 0.17 & $-$6.64 & 0.23 \\
  (10, 100) & $-$3.88 & 0.10 & $-$4.50 & 0.13 \\
  (5, 500) & $-$6.42 & 0.06 & $-$1.05 & 0.04 \\
  (10, 500) & $-$2.97 & 0.03 & $-$1.33 & 0.02 \\
   \bottomrule
\end{tabular}
\label{tab:B.9}
\end{table}

\clearpage

\subsection{Heteroscedastic scenario}

\begin{table}[ht]
\caption{Quantile 0.1 and heteroscedastic scenario. Average bias and root mean squared error (RMSE) for $Q(0.1)$ from the additive quantile mixed model (AQMM) and the additive fixed-effects quantile regression (AFEQR). The expected proportion of negative residuals (PNR) is 0.1.}
\centering
\begin{tabular}{lrrrrrr}
  \toprule
Sample size & \multicolumn{3}{c}{AQMM} & \multicolumn{3}{c}{AFEQR} \\
($n$, $M$) & PNR & Bias & RMSE & PNR & Bias & RMSE \\
  \midrule
  \multicolumn{7}{l}{\textit{Normal}}\\
  \midrule
(5, 50) & 0.11 & $-$0.15 & 1.69 & 0.10 & $-$2.02 & 3.13 \\
  (10, 50) & 0.10 & $-$0.25 & 1.76 & 0.10 & $-$2.16 & 3.15 \\
  (5, 100) & 0.10 & $-$0.15 & 1.67 & 0.10 & $-$2.17 & 3.17 \\
  (10, 100) & 0.10 & $-$0.25 & 1.74 & 0.10 & $-$2.22 & 3.13 \\
  (5, 500) & 0.10 & $-$0.17 & 1.64 & 0.10 & $-$2.19 & 3.01 \\
  (10, 500) & 0.10 & $-$0.26 & 1.72 & 0.10 & $-$2.19 & 3.00 \\
  \midrule
  \multicolumn{7}{l}{\textit{Student's $t$}}\\
  \midrule
  (5, 50) & 0.10 & $-$0.63 & 1.85 & 0.10 & $-$2.00 & 3.22 \\
  (10, 50) & 0.10 & $-$0.61 & 1.83 & 0.10 & $-$2.10 & 3.19 \\
  (5, 100) & 0.10 & $-$0.54 & 1.73 & 0.10 & $-$2.13 & 3.23 \\
  (10, 100) & 0.10 & $-$0.55 & 1.77 & 0.10 & $-$2.17 & 3.16 \\
  (5, 500) & 0.10 & $-$0.54 & 1.62 & 0.10 & $-$2.14 & 3.02 \\
  (10, 500) & 0.10 & $-$0.54 & 1.74 & 0.10 & $-$2.14 & 3.01 \\
  \midrule
  \multicolumn{7}{l}{\textit{Chi-squared}}\\
  \midrule
  (5, 50) & 0.09 & 0.02 & 1.72 & 0.10 & $-$1.62 & 2.94 \\
  (10, 50) & 0.09 & $-$0.01 & 1.63 & 0.10 & $-$1.77 & 2.93 \\
  (5, 100) & 0.09 & 0.01 & 1.56 & 0.10 & $-$1.80 & 2.97 \\
  (10, 100) & 0.09 & $-$0.03 & 1.55 & 0.10 & $-$1.86 & 2.88 \\
  (5, 500) & 0.09 & 0.06 & 1.38 & 0.10 & $-$1.85 & 2.78 \\
  (10, 500) & 0.10 & 0.13 & 1.47 & 0.10 & $-$1.85 & 2.77 \\
   \bottomrule
\end{tabular}
\label{tab:B.10}
\end{table}

\begin{table}[ht]
\caption{Quantile 0.5 and heteroscedastic scenario. Average bias and root mean squared error (RMSE) for $Q(0.5)$ from the additive quantile mixed model (AQMM), the additive fixed-effects quantile regression (AFEQR), and the additive mixed model (AMM). The expected proportion of negative residuals (PNR) is 0.5.}
\centering
\begin{tabular}{lrrrrrrrrr}
  \toprule
Sample size & \multicolumn{3}{c}{AQMM} & \multicolumn{3}{c}{AFEQR} & \multicolumn{3}{c}{AMM}\\
($n$, $M$) & PNR & Bias & RMSE & PNR & Bias & RMSE & PNR & Bias & RMSE\\
  \midrule
  \multicolumn{10}{l}{\textit{Normal}}\\
  \midrule
(5, 50) & 0.50 & $-$0.01 & 1.62 & 0.50 & $-$0.01 & 1.77 & 0.50 & $-$0.01 & 1.61 \\
  (10, 50) & 0.50 & $-$0.00 & 1.66 & 0.50 & $-$0.02 & 1.57 & 0.50 & $-$0.01 & 1.66 \\
  (5, 100) & 0.50 & 0.01 & 1.61 & 0.50 & $-$0.00 & 1.58 & 0.50 & 0.01 & 1.60 \\
  (10, 100) & 0.50 & $-$0.00 & 1.65 & 0.50 & $-$0.02 & 1.42 & 0.50 & 0.00 & 1.65 \\
  (5, 500) & 0.50 & 0.00 & 1.59 & 0.50 & $-$0.02 & 1.30 & 0.50 & 0.00 & 1.59 \\
  (10, 500) & 0.50 & 0.00 & 1.64 & 0.50 & $-$0.02 & 1.29 & 0.50 & 0.00 & 1.64 \\
  \midrule
  \multicolumn{10}{l}{\textit{Student's $t$}}\\
  \midrule
  (5, 50) & 0.50 & $-$0.01 & 1.56 & 0.50 & $-$0.01 & 1.89 & 0.50 & $-$0.01 & 1.58 \\
  (10, 50) & 0.50 & $-$0.01 & 1.56 & 0.50 & $-$0.04 & 1.65 & 0.50 & $-$0.01 & 1.58 \\
  (5, 100) & 0.50 & 0.00 & 1.50 & 0.50 & $-$0.01 & 1.67 & 0.50 & $-$0.01 & 1.52 \\
  (10, 100) & 0.50 & 0.00 & 1.54 & 0.50 & $-$0.02 & 1.51 & 0.50 & 0.00 & 1.55 \\
  (5, 500) & 0.50 & 0.00 & 1.42 & 0.50 & $-$0.02 & 1.36 & 0.50 & 0.00 & 1.43 \\
  (10, 500) & 0.50 & 0.00 & 1.53 & 0.50 & $-$0.03 & 1.36 & 0.50 & $-$0.00 & 1.53 \\
  \midrule
  \multicolumn{10}{l}{\textit{Chi-squared}}\\
  \midrule
  (5, 50) & 0.50 & 1.06 & 2.23 & 0.50 & 0.40 & 2.19 & 0.83 & 3.18 & 3.55 \\
  (10, 50) & 0.50 & 0.99 & 2.13 & 0.50 & 0.36 & 1.92 & 0.83 & 3.20 & 3.56 \\
  (5, 100) & 0.50 & 1.04 & 2.10 & 0.50 & 0.36 & 1.97 & 0.83 & 3.18 & 3.50 \\
  (10, 100) & 0.50 & 0.95 & 2.05 & 0.50 & 0.32 & 1.75 & 0.83 & 3.18 & 3.51 \\
  (5, 500) & 0.50 & 0.98 & 1.95 & 0.50 & 0.30 & 1.56 & 0.83 & 3.19 & 3.44 \\
  (10, 500) & 0.50 & 0.92 & 1.99 & 0.50 & 0.30 & 1.55 & 0.83 & 3.19 & 3.49 \\
   \bottomrule
\end{tabular}
\label{tab:B.11}
\end{table}

\begin{table}[ht]
\caption{Quantile 0.95 and heteroscedastic scenario. Average bias and root mean squared error (RMSE) for $Q(0.95)$ from the additive quantile mixed model (AQMM) and the additive fixed-effects quantile regression (AFEQR). The expected proportion of negative residuals (PNR) is 0.95.}
\centering
\begin{tabular}{lrrrrrr}
  \toprule
Sample size & \multicolumn{3}{c}{AQMM} & \multicolumn{3}{c}{AFEQR} \\
($n$, $M$) & PNR & Bias & RMSE & PNR & Bias & RMSE \\
  \midrule
  \multicolumn{7}{l}{\textit{Normal}}\\
  \midrule
(5, 50) & 0.94 & 0.18 & 1.73 & 0.95 & 2.79 & 3.90 \\
  (10, 50) & 0.95 & 0.31 & 1.81 & 0.95 & 2.97 & 3.99 \\
  (5, 100) & 0.95 & 0.22 & 1.72 & 0.95 & 3.02 & 4.04 \\
  (10, 100) & 0.95 & 0.33 & 1.80 & 0.95 & 3.08 & 4.03 \\
  (5, 500) & 0.95 & 0.23 & 1.68 & 0.95 & 3.06 & 3.94 \\
  (10, 500) & 0.95 & 0.35 & 1.78 & 0.95 & 3.06 & 3.94 \\
  \midrule
  \multicolumn{7}{l}{\textit{Student's $t$}}\\
  \midrule
  (5, 50) & 0.95 & 0.72 & 2.07 & 0.95 & 2.48 & 3.82 \\
  (10, 50) & 0.95 & 0.74 & 2.06 & 0.95 & 2.64 & 3.85 \\
  (5, 100) & 0.95 & 0.65 & 1.92 & 0.95 & 2.69 & 3.89 \\
  (10, 100) & 0.95 & 0.71 & 1.98 & 0.95 & 2.76 & 3.87 \\
  (5, 500) & 0.95 & 0.62 & 1.77 & 0.95 & 2.71 & 3.74 \\
  (10, 500) & 0.95 & 0.66 & 1.89 & 0.95 & 2.72 & 3.73 \\
  \midrule
  \multicolumn{7}{l}{\textit{Chi-squared}}\\
  \midrule
  (5, 50) & 0.95 & 1.73 & 4.08 & 0.95 & 1.15 & 3.73 \\
  (10, 50) & 0.95 & 1.99 & 4.26 & 0.95 & 1.50 & 3.60 \\
  (5, 100) & 0.95 & 1.80 & 4.07 & 0.95 & 1.53 & 3.63 \\
  (10, 100) & 0.95 & 2.00 & 4.24 & 0.95 & 1.64 & 3.48 \\
  (5, 500) & 0.95 & 1.82 & 3.95 & 0.95 & 1.57 & 3.24 \\
  (10, 500) & 0.95 & 1.96 & 4.11 & 0.95 & 1.58 & 3.21 \\
   \bottomrule
\end{tabular}
\label{tab:B.12}
\end{table}

\begin{table}[ht]
\caption{Quantile 0.1 and heteroscedastic scenario. Average relative bias (\%) and root mean squared error (RMSE) for $\beta_3$ from the additive quantile mixed model (AQMM) and the additive fixed-effects quantile regression (AFEQR).}
\centering
\begin{tabular}{lrrrr}
  \toprule
Sample size & \multicolumn{2}{c}{AQMM} & \multicolumn{2}{c}{AFEQR} \\
($n$, $M$) & Rel. bias & RMSE & Rel. bias & RMSE \\
  \midrule
  \multicolumn{5}{l}{\textit{Normal}}\\
  \midrule
(5, 50) & $-$0.43 & 0.69 & $-$7.12 & 0.54 \\
  (10, 50) & $-$0.48 & 0.43 & $-$9.88 & 0.35 \\
  (5, 100) & $-$0.29 & 0.29 & $-$9.48 & 0.35 \\
  (10, 100) & $-$3.19 & 0.14 & $-$9.85 & 0.23 \\
  (5, 500) & $-$4.14 & 0.07 & $-$10.40 & 0.17 \\
  (10, 500) & $-$4.53 & 0.05 & $-$10.45 & 0.15 \\
  \midrule
  \multicolumn{5}{l}{\textit{Student's $t$}}\\
  \midrule
  (5, 50) & 3.52 & 0.95 & $-$10.17 & 0.75 \\
  (10, 50) & $-$0.46 & 0.56 & $-$10.82 & 0.44 \\
  (5, 100) & $-$2.96 & 0.48 & $-$10.93 & 0.46 \\
  (10, 100) & $-$4.46 & 0.23 & $-$10.78 & 0.27 \\
  (5, 500) & $-$5.27 & 0.11 & $-$11.71 & 0.20 \\
  (10, 500) & $-$6.34 & 0.08 & $-$12.47 & 0.18 \\
  \midrule
  \multicolumn{5}{l}{\textit{Chi-squared}}\\
  \midrule
  (5, 50) & 3.88 & 0.85 & $-$7.07 & 0.68 \\
  (10, 50) & 1.12 & 0.45 & $-$9.09 & 0.54 \\
  (5, 100) & 0.55 & 0.34 & $-$9.37 & 0.52 \\
  (10, 100) & $-$1.97 & 0.15 & $-$9.97 & 0.42 \\
  (5, 500) & $-$2.65 & 0.07 & $-$9.39 & 0.30 \\
  (10, 500) & $-$2.67 & 0.05 & $-$10.28 & 0.33 \\
   \bottomrule
\end{tabular}
\label{tab:B.13}
\end{table}

\begin{table}[ht]
\caption{Quantile 0.5 and heteroscedastic scenario. Average relative bias (\%) and root mean squared error (RMSE) for $\beta_3$ from the additive quantile mixed model (AQMM), the additive fixed-effects quantile regression (AFEQR), and the additive mixed model (AMM).}
\centering
\begin{tabular}{lrrrrrr}
  \toprule
Sample size & \multicolumn{2}{c}{AQMM} & \multicolumn{2}{c}{AFEQR} & \multicolumn{2}{c}{AMM}\\
($n$, $M$) & Rel. bias & RMSE & Rel. bias & RMSE & Rel. bias & RMSE\\
  \midrule
  \multicolumn{7}{l}{\textit{Normal}}\\
  \midrule
(5, 50) & $-$0.73 & 0.10 & $-$5.61 & 0.33 & $-$0.85 & 0.22 \\
  (10, 50) & $-$0.20 & 0.05 & $-$5.88 & 0.19 & $-$0.16 & 0.15 \\
  (5, 100) & 0.17 & 0.05 & $-$5.21 & 0.17 & 0.23 & 0.16 \\
  (10, 100) & $-$0.02 & 0.02 & $-$5.31 & 0.10 & 0.03 & 0.11 \\
  (5, 500) & $-$0.11 & 0.01 & $-$5.12 & 0.07 & $-$0.10 & 0.07 \\
  (10, 500) & 0.06 & 0.00 & $-$5.34 & 0.06 & 0.02 & 0.05 \\
  \midrule
  \multicolumn{7}{l}{\textit{Student's $t$}}\\
  \midrule
  (5, 50) & $-$0.60 & 0.17 & $-$5.10 & 0.36 & $-$0.86 & 0.36 \\
  (10, 50) & $-$0.29 & 0.07 & $-$7.07 & 0.22 & $-$0.44 & 0.24 \\
  (5, 100) & $-$0.46 & 0.08 & $-$5.68 & 0.21 & $-$1.04 & 0.23 \\
  (10, 100) & 0.07 & 0.03 & $-$5.96 & 0.13 & 0.30 & 0.16 \\
  (5, 500) & $-$0.15 & 0.02 & $-$5.53 & 0.08 & $-$0.07 & 0.11 \\
  (10, 500) & $-$0.06 & 0.01 & $-$6.10 & 0.07 & $-$0.16 & 0.08 \\
  \midrule
  \multicolumn{7}{l}{\textit{Chi-squared}}\\
  \midrule
  (5, 50) & 0.88 & 0.43 & $-$4.49 & 0.59 & 9.25 & 0.69 \\
  (10, 50) & $-$0.20 & 0.22 & $-$6.09 & 0.43 & 9.47 & 0.63 \\
  (5, 100) & 0.71 & 0.21 & $-$5.66 & 0.36 & 9.60 & 0.65 \\
  (10, 100) & $-$0.14 & 0.10 & $-$6.14 & 0.28 & 9.48 & 0.61 \\
  (5, 500) & 0.62 & 0.04 & $-$5.47 & 0.17 & 10.00 & 0.64 \\
  (10, 500) & 0.12 & 0.02 & $-$5.53 & 0.15 & 9.83 & 0.63 \\
   \bottomrule
\end{tabular}
\label{tab:B.14}
\end{table}

\begin{table}[ht]
\caption{Quantile 0.95 and heteroscedastic scenario. Average relative bias (\%) and root mean squared error (RMSE) for $\beta_3$ from the additive quantile mixed model (AQMM) and the additive fixed-effects quantile regression (AFEQR).}
\centering
\begin{tabular}{lrrrr}
  \toprule
Sample size & \multicolumn{2}{c}{AQMM} & \multicolumn{2}{c}{AFEQR} \\
($n$, $M$) & Rel. bias & RMSE & Rel. bias & RMSE \\
  \midrule
  \multicolumn{5}{l}{\textit{Normal}}\\
  \midrule
(5, 50) & $-$4.48 & 1.11 & $-$10.65 & 0.91 \\
  (10, 50) & $-$2.10 & 0.57 & $-$10.28 & 0.58 \\
  (5, 100) & $-$0.43 & 0.50 & $-$10.30 & 0.59 \\
  (10, 100) & 0.96 & 0.23 & $-$8.59 & 0.34 \\
  (5, 500) & 2.24 & 0.09 & $-$7.87 & 0.21 \\
  (10, 500) & 1.83 & 0.05 & $-$7.66 & 0.17 \\
  \midrule
  \multicolumn{5}{l}{\textit{Student's $t$}}\\
  \midrule
  (5, 50) & $-$1.36 & 1.99 & $-$12.09 & 1.56 \\
  (10, 50) & $-$1.26 & 0.96 & $-$11.89 & 0.90 \\
  (5, 100) & 0.23 & 0.77 & $-$12.67 & 0.97 \\
  (10, 100) & 1.48 & 0.48 & $-$11.12 & 0.64 \\
  (5, 500) & 3.07 & 0.18 & $-$9.86 & 0.40 \\
  (10, 500) & 3.00 & 0.11 & $-$9.68 & 0.32 \\
  \midrule
  \multicolumn{5}{l}{\textit{Chi-squared}}\\
  \midrule
  (5, 50) & $-$4.72 & 5.62 & $-$20.24 & 8.11 \\
  (10, 50) & $-$4.21 & 3.14 & $-$16.53 & 5.23 \\
  (5, 100) & $-$1.14 & 2.77 & $-$16.28 & 5.38 \\
  (10, 100) & 1.24 & 1.60 & $-$13.81 & 3.41 \\
  (5, 500) & 2.20 & 0.63 & $-$11.84 & 2.16 \\
  (10, 500) & 1.86 & 0.31 & $-$11.44 & 1.79 \\
   \bottomrule
\end{tabular}
\label{tab:B.15}
\end{table}

\begin{table}[ht]
\caption{Quantile 0.1 and heteroscedastic scenario. Average relative bias (\%) and root mean squared error (RMSE) for $\beta_4$ from the additive quantile mixed model (AQMM) and the additive fixed-effects quantile regression (AFEQR).}
\centering
\begin{tabular}{lrrrr}
  \toprule
Sample size & \multicolumn{2}{c}{AQMM} & \multicolumn{2}{c}{AFEQR} \\
($n$, $M$) & Rel. bias & RMSE & Rel. bias & RMSE \\
  \midrule
  \multicolumn{5}{l}{\textit{Normal}}\\
  \midrule
(5, 50) & 1.02 & 0.04 & $-$16.29 & 0.34 \\
  (10, 50) & $-$0.04 & 0.03 & $-$16.38 & 0.31 \\
  (5, 100) & 0.92 & 0.02 & $-$16.24 & 0.30 \\
  (10, 100) & 0.24 & 0.01 & $-$16.16 & 0.27 \\
  (5, 500) & 1.23 & 0.01 & $-$15.97 & 0.24 \\
  (10, 500) & 0.35 & 0.00 & $-$16.31 & 0.25 \\
  \midrule
  \multicolumn{5}{l}{\textit{Student's $t$}}\\
  \midrule
  (5, 50) & 1.44 & 0.08 & $-$17.81 & 0.44 \\
  (10, 50) & 0.77 & 0.04 & $-$16.87 & 0.32 \\
  (5, 100) & 1.17 & 0.04 & $-$17.22 & 0.34 \\
  (10, 100) & $-$0.17 & 0.02 & $-$16.90 & 0.30 \\
  (5, 500) & 1.48 & 0.01 & $-$16.59 & 0.26 \\
  (10, 500) & 0.47 & 0.00 & $-$16.74 & 0.26 \\
  \midrule
  \multicolumn{5}{l}{\textit{Chi-squared}}\\
  \midrule
  (5, 50) & 1.40 & 0.11 & $-$14.22 & 0.32 \\
  (10, 50) & 0.58 & 0.05 & $-$13.90 & 0.25 \\
  (5, 100) & 1.13 & 0.04 & $-$14.35 & 0.25 \\
  (10, 100) & 0.28 & 0.02 & $-$14.62 & 0.23 \\
  (5, 500) & 0.77 & 0.01 & $-$14.20 & 0.20 \\
  (10, 500) & 0.10 & 0.00 & $-$14.52 & 0.20 \\
   \bottomrule
\end{tabular}
\label{tab:B.16}
\end{table}

\begin{table}[ht]
\caption{Quantile 0.5 and heteroscedastic scenario. Average relative bias (\%) and root mean squared error (RMSE) for $\beta_4$ from the additive quantile mixed model (AQMM), the additive fixed-effects quantile regression (AFEQR), and the additive mixed model (AMM).}
\centering
\begin{tabular}{lrrrrrr}
  \toprule
 Sample size  & \multicolumn{2}{c}{AQMM} & \multicolumn{2}{c}{AFEQR} & \multicolumn{2}{c}{AMM}\\
   ($n$, $M$) & Rel. bias & RMSE & Rel. bias & RMSE & Rel. bias & RMSE\\
  \midrule
  \multicolumn{7}{l}{\textit{Normal}}\\
  \midrule
(5, 50) & $-$0.33 & 0.03 & $-$5.83 & 0.10 & $-$0.30 & 0.14 \\
  (10, 50) & $-$0.27 & 0.03 & $-$6.77 & 0.09 & $-$0.38 & 0.13 \\
  (5, 100) & $-$0.13 & 0.02 & $-$5.44 & 0.06 & $-$0.21 & 0.10 \\
  (10, 100) & $-$0.09 & 0.01 & $-$5.77 & 0.05 & $-$0.05 & 0.09 \\
  (5, 500) & 0.09 & 0.00 & $-$5.40 & 0.03 & 0.10 & 0.04 \\
  (10, 500) & 0.02 & 0.00 & $-$5.56 & 0.03 & 0.01 & 0.04 \\
  \midrule
  \multicolumn{7}{l}{\textit{Student's $t$}}\\
  \midrule
  (5, 50) & $-$0.15 & 0.05 & $-$6.01 & 0.11 & $-$0.09 & 0.18 \\
  (10, 50) & $-$0.10 & 0.03 & $-$6.53 & 0.08 & $-$0.10 & 0.14 \\
  (5, 100) & $-$0.18 & 0.02 & $-$5.36 & 0.06 & 0.01 & 0.13 \\
  (10, 100) & $-$0.12 & 0.02 & $-$6.25 & 0.06 & $-$0.17 & 0.10 \\
  (5, 500) & 0.11 & 0.00 & $-$5.49 & 0.03 & 0.08 & 0.06 \\
  (10, 500) & 0.08 & 0.00 & $-$5.64 & 0.03 & 0.10 & 0.05 \\
  \midrule
  \multicolumn{7}{l}{\textit{Chi-squared}}\\
  \midrule
  (5, 50) & 2.03 & 0.10 & $-$5.38 & 0.13 & $-$1.06 & 0.22 \\
  (10, 50) & 1.12 & 0.05 & $-$5.88 & 0.09 & $-$0.32 & 0.18 \\
  (5, 100) & 1.97 & 0.04 & $-$4.77 & 0.07 & $-$0.26 & 0.15 \\
  (10, 100) & 0.92 & 0.02 & $-$5.03 & 0.05 & $-$0.45 & 0.12 \\
  (5, 500) & 1.99 & 0.01 & $-$4.41 & 0.03 & 0.02 & 0.07 \\
  (10, 500) & 1.44 & 0.01 & $-$4.63 & 0.03 & 0.03 & 0.05 \\
   \bottomrule
\end{tabular}
\label{tab:B.17}
\end{table}

\begin{table}[ht]
\caption{Quantile 0.95 and heteroscedastic scenario. Average relative bias (\%) and root mean squared error (RMSE) for $\beta_4$ from the additive quantile mixed model (AQMM) and the additive fixed-effects quantile regression (AFEQR).}
\centering
\begin{tabular}{lrrrr}
  \toprule
Sample size & \multicolumn{2}{c}{AQMM} & \multicolumn{2}{c}{AFEQR} \\
($n$, $M$) & Rel. bias & RMSE & Rel. bias & RMSE \\
  \midrule
  \multicolumn{5}{l}{\textit{Normal}}\\
  \midrule
(5, 50) & $-$1.70 & 0.05 & $-$0.32 & 0.19 \\
  (10, 50) & $-$0.75 & 0.04 & 2.33 & 0.10 \\
  (5, 100) & $-$1.78 & 0.03 & 3.95 & 0.09 \\
  (10, 100) & $-$0.20 & 0.02 & 4.35 & 0.07 \\
  (5, 500) & $-$1.15 & 0.01 & 5.15 & 0.04 \\
  (10, 500) & $-$0.39 & 0.00 & 5.26 & 0.03 \\
  \midrule
  \multicolumn{5}{l}{\textit{Student's $t$}}\\
  \midrule
  (5, 50) & $-$2.62 & 0.13 & $-$1.04 & 0.27 \\
  (10, 50) & $-$1.35 & 0.08 & 0.32 & 0.14 \\
  (5, 100) & $-$1.55 & 0.06 & 1.76 & 0.12 \\
  (10, 100) & $-$0.79 & 0.04 & 2.47 & 0.08 \\
  (5, 500) & $-$1.57 & 0.01 & 4.74 & 0.04 \\
  (10, 500) & $-$0.61 & 0.01 & 4.59 & 0.03 \\
  \midrule
  \multicolumn{5}{l}{\textit{Chi-squared}}\\
  \midrule
  (5, 50) & $-$5.67 & 0.39 & $-$5.62 & 0.51 \\
  (10, 50) & $-$3.28 & 0.22 & $-$1.00 & 0.26 \\
  (5, 100) & $-$5.20 & 0.20 & $-$1.46 & 0.26 \\
  (10, 100) & $-$4.23 & 0.13 & 0.10 & 0.13 \\
  (5, 500) & $-$5.78 & 0.06 & 3.16 & 0.06 \\
  (10, 500) & $-$3.23 & 0.03 & 3.11 & 0.04 \\
   \bottomrule
\end{tabular}
\label{tab:B.18}
\end{table}

%\bibliographystyle{imsart-nameyear}
%\bibliography{biblio}

\end{document}